\pdfoutput=1

\documentclass[11pt]{article}

\usepackage[final]{acl}

\usepackage{times}
\usepackage{latexsym}

\usepackage[T1]{fontenc}

\usepackage[utf8]{inputenc}

\usepackage{microtype}

\usepackage{inconsolata}

\usepackage{graphicx}

\usepackage{microtype}
\usepackage{graphicx}
\usepackage{subfigure}
\usepackage{booktabs} 
\usepackage{hyperref}
\usepackage{amsmath}
\usepackage{amssymb}
\usepackage{mathtools}
\usepackage{amsthm}

\usepackage[textsize=tiny]{todonotes}
\usepackage{xspace}

\usepackage{adjustbox}
\usepackage[frozencache,cachedir=minted-cache]{minted}
\definecolor{pastel_blue}{rgb}{0.86, 0.926, 0.984}

\usepackage{graphicx}
\usepackage{listings}
\usepackage{caption}
\usepackage{amsmath}
\usepackage{tcolorbox}
\usepackage[table]{xcolor} 
\definecolor{pos}{RGB}{210,245,239}
\definecolor{neg}{RGB}{253,213,210}

\newcommand{\Coder}{ExeCoder\xspace}

\title{\Coder: Empowering Large Language Models with Executability Representation for Code Translation}

\author{Minghua He\thanks{These authors contributed equally to this work.}
\thanks{This work was done during the internship at Microsoft.}$^{1}$, Yue Chen\footnotemark[1]\footnotemark[2]$^{1}$, Fangkai Yang\thanks{Corresponding author.}$^{2}$, Pu Zhao$^{2}$, Wenjie Yin$^{3}$, \\ \textbf{Yu Kang$^{2}$, Qingwei Lin$^{2}$, Saravan Rajmohan$^{2}$, Dongmei Zhang$^{2}$}  \\
  $^{1}$Peking University,
  $^{2}$Microsoft,
  $^{3}$KTH Royal Institute of Technology\\
\texttt{hemh2120@stu.pku.edu.cn, fangkai.yang@microsoft.com}
}

\begin{document}
\maketitle


\begin{abstract}
Code translation is a crucial activity in the software development and maintenance process, and researchers have recently begun to focus on using pre-trained large language models~(LLMs) for code translation. However, existing LLMs only learn the contextual semantics of code during pre-training, neglecting executability information closely related to the execution state of the code, which results in unguaranteed code executability and unreliable automated code translation. To address this issue, we propose \Coder, an LLM specifically designed for code translation, aimed at utilizing executability representations such as functional semantics, syntax structures, and variable dependencies to enhance the capabilities of LLMs in code translation. To evaluate the effectiveness of \Coder, we manually enhanced the widely used benchmark TransCoder-test, resulting in a benchmark called TransCoder-test-X that serves LLMs. Evaluation of TransCoder-test-X indicates that \Coder achieves state-of-the-art performance in code translation, surpassing existing open-source code LLMs by over 10.88\% to 38.78\% and over 27.44\% to 42.97\% on two metrics, and even outperforms the renowned closed-source LLM GPT-4o. Code is available at \url{https://aka.ms/execoder}
\end{abstract}

\section{Introduction}

Code translation aims to convert code written in one programming language into another. Translation between different languages can assist developers in adapting applications to new business and environments, demonstrating significant demand and value in real industrial contexts. For example, the Commonwealth Bank of Australia spent around \$750 million and five years converting its platform from COBOL to Java \cite{transcoder}. 


\lstset{
  escapeinside={(*@}{@*)}  
}

\lstdefinestyle{lfonts}{
  backgroundcolor=\color{white!5},
  basicstyle   = \footnotesize\ttfamily,
  stringstyle  = \color{purple},
  keywordstyle = \color{green!60!black}\bfseries,
  commentstyle = \color{olive},
}
\lstdefinestyle{lnumbers}{
}
\lstdefinestyle{llayout}{
  breaklines       = true,
  tabsize          = 2,
  columns          = flexible,
}
\lstdefinestyle{lgeometry}{
}
\lstdefinestyle{lgeneral}{
  style = lfonts,
}
\lstdefinestyle{python}{
  language = {Python},
  style    = lgeneral,
}

\begin{figure}[t]
\centering
\begin{adjustbox}{width=0.9\linewidth,center}
\begin{tabular}{l l l}

\midrule
\footnotesize{Input Function}\\
\midrule
\begin{lstlisting}[style=Python, xleftmargin=0cm, basicstyle=\fontencoding{T1}\fontfamily{lmtt}\scriptsize]
# C++
int findMaxAverage(int arr[], int n, int k) {
    if (k > n) return -1;
    int sum = arr[0];
    for (int i = 1; i < k; i++) sum += arr[i];
    int max_sum = sum, max_end = k - 1;
}
\end{lstlisting} \\

\midrule
\footnotesize{Translation w/o Executability Representation}\\
\midrule
\begin{lstlisting}[style=Python, xleftmargin=0cm, basicstyle=\fontencoding{T1}\fontfamily{lmtt}\scriptsize]
# Python
def findMaxAverage(arr, n, k):
    if k > n:
        return -1
    sum = sum(arr[:k])
    max_sum = sum
    max_end = k - 1
\end{lstlisting} \\

\midrule
\footnotesize{Translation with Executability Representation}\\
\midrule
\begin{lstlisting}[style=Python, xleftmargin=0cm, basicstyle=\fontencoding{T1}\fontfamily{lmtt}\scriptsize]
# Python
def findMaxAverage ( arr , n , k ) :
    if ( k > n ) : return - 1
    sum = arr [ 0 ]
    for i in range ( 1 , k ) : sum += arr [ i ]
    max_sum = sum
    max_end = k - 1
\end{lstlisting}\\

 \end{tabular}
 \end{adjustbox}
    \caption{\textbf{Executability Representation for Code Translation.} Existing models simply copy variable name from the source code, ignoring conflicts with called built-in function, which leads to TypeError exception. The model that considers code executability learn the syntactic structure of source code and avoid call conflicts through explicit loop summation.}
    
\vspace{-0.6cm}
\label{fig:teaser}
\end{figure}

Given the high labor costs associated with code translation, the development of automated translation tools has been extensively researched. Recent approaches \cite{huang2023program, roziereleveraging, wen2022babeltower, zhu2022multilingual} utilize the paradigm of neural machine translation (NMT), aiming to learn the mapping from source language sequences to target language sequences. Despite achieving some success, their effectiveness heavily relies on the availability of cross-language parallel data \cite{ahmad2021avatar}. For most languages, parallel resources are scarce or entirely absent. To overcome the limitations of NMT-based approaches, some studies \cite{yin2024rectifier, lu2025axisefficienthumanagentcomputerinteraction, mei2025a1steeptesttimescaling, macedo2024exploring, yang2024exploring} have explored the use of large language models (LLMs) for code translation. These LLMs are pre-trained on a vast array of open-source code, generating code by learning the contextual semantics of the code and demonstrating excellent understanding across various programming languages. Nevertheless, existing research indicates that the correct translation rate of advanced LLMs ranges from 2.1\% to 47.3\%, resulting in 15 different types of execution errors \cite{pan2024lost}. Leveraging the potential of LLMs for code translation poses significant challenges.

In our view, the fundamental reason for this issue is the discrepancy between the pre-training tasks of existing LLMs and the requirements of code translation, as illustrated in Figure \ref{fig:teaser}. Unlike natural languages, programming languages possess additional information that indicates the execution state of the code, including more complex functionalities, syntax, and variables, referred to as executability information \cite{jiao2023evaluation}. Executability information pertains to the implementation logic of the source code, the acquisition of which relies on code analysis tools designed by programming language experts and cannot be directly inferred from the context of the source code. Code translation requires that the generated code executes correctly as intended. However, existing LLMs only learn the contextual semantics of code during pre-training, neglecting the executability information closely related to the execution state of the code, which compromises the executability of the generated code and prevents reliable automated code translation.

To address these issues and harness the potential of LLMs in code translation, we propose ExeCoder. This is an LLM specifically designed for code translation, aimed at enhancing the ability of LLMs by leveraging executability representations such as functional semantics, syntactic structures, and variable dependencies within the code. To extract executability knowledge from the code, ExeCoder first devises a representation strategy for executability knowledge. ExeCoder employs external code analysis tools to acquire three types of executability knowledge from the source code and meticulously designs encoding strategies to convert this knowledge into text that is easily interpretable by LLMs. To learn executability knowledge from the code, ExeCoder introduces a Progressive Executability Representation Learning strategy (PERL). The key idea is that the functional semantics, syntactic structures, and variable dependencies of the source code represent progressively refined executability information, and this staged, progressive learning aligns with the learning theories of programming experts, facilitating representation learning.

To evaluate the effectiveness of ExeCoder, we conducted evaluation on the widely used code translation benchmark, TransCoder-test \cite{transcoder}. However, TransCoder-test can only evaluate specific implementations of the code. To address this limitation, we enhanced the TransCoder-test benchmark, resulting in a new benchmark called TransCoder-test-X, capable of evaluating the code translation capabilities of LLMs. Evaluation on TransCoder-test-X indicates that ExeCoder achieves SOTA performance in code translation tasks, surpassing existing open-source large models by over 10.88\% to 38.78\% and over 27.44\% to 42.97\% on two metrics, respectively. Notably, the ExeCoder outperforms renowned closed-source LLMs, including GPT-3.5, GPT-4, and GPT-4o, highlighting the significant role of executability representations in code translation.

In summary, our contributions are as follows: 
\begin{itemize}
\item We developed ExeCoder, a LLM specifically designed for code translation, which significantly outperforms all other open-source code LLMs, achieving SOTA performance. Notably, the ExeCoder surpasses well-known the renowned closed-source LLM GPT-4o.
\item We propose a Progressive Executability Representation Learning strategy that aligns with the learning theory of programming experts and effectively learns executability representations of code.
\item We enhanced the widely used code translation benchmark, TransCoder-test, resulting in a new benchmark called TransCoder-test-X, which is capable of evaluating the code translation abilities of LLMs.
\item We conducted a preliminary study that emphasizes the critical role of executability representations of code in achieving excellent code translation performance.
\end{itemize}

 \begin{figure*}[t]
  \centering
  \includegraphics[width=0.9\linewidth]{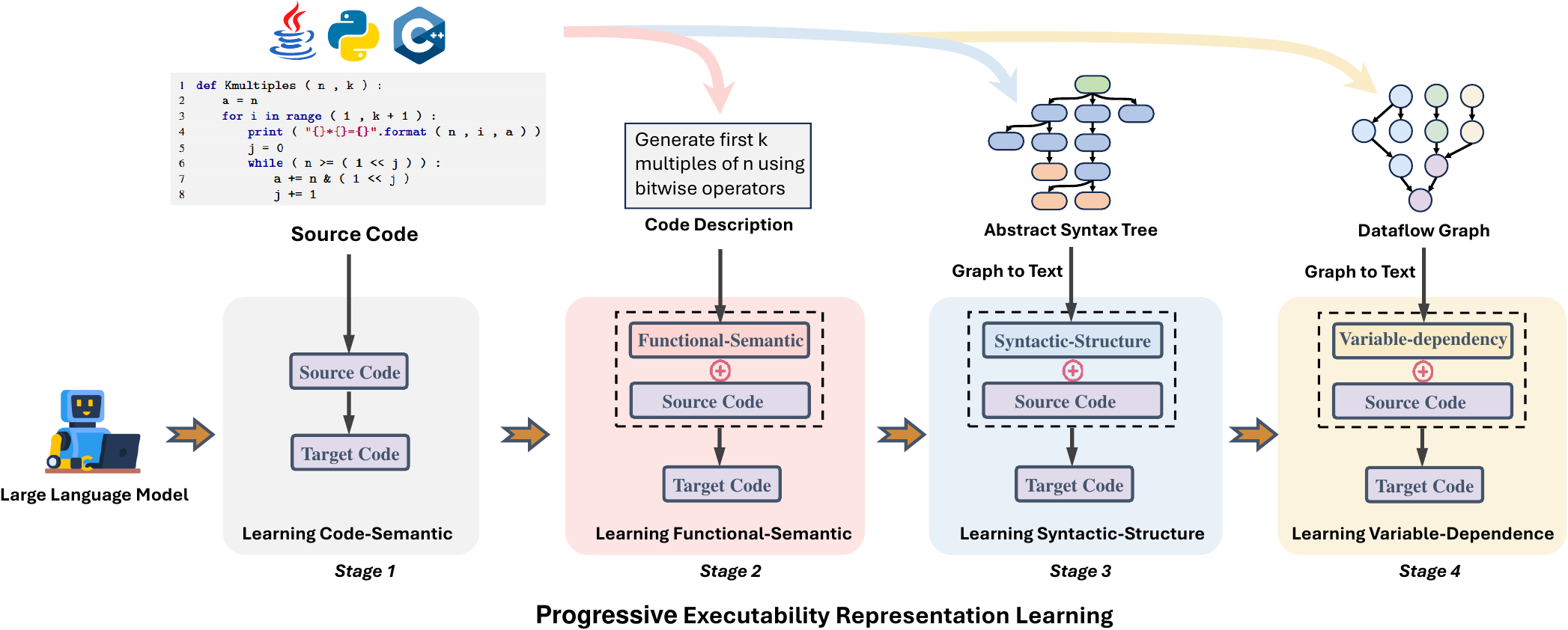}
  \caption{\textbf{The \Coder's pipeline.} The \Coder first utilizes a text that describes the code functionality to encode functional semantic information. Subsequently, the \Coder uses an abstract syntax tree (AST) to encode the syntactic structure information. Finally, the \Coder employs a data flow graph (DFG) to encode the variable dependency information. With the benefit of the well-designed Progressive Executability Representation Learning (PERL) strategy, the \Coder can fully leverage the executability representation to enhance the cross-language understanding capabilities of LLMs.}
  \label{fig:overview}
\vspace{-0.5cm}
\end{figure*}

\section{Related Work}
\subsection{Translation of Programming Languages}
The translation between programming languages has been an active research field \cite{liu2024hmcodetrans, eniser2024automatically, khan2024xcodeeval, zhu2024semi, luo2024bridging}, recent work has been based on the paradigm of neural machine translation. 
TransCoder \cite{transcoder} is the most classic unsupervised machine translation method, specifically designed with pre-training tasks to learn the semantics of source code. 
Other works aim to leverage code analysis tools to enhance the capabilities of code translation. 
SDA-Trans \cite{liu2023syntax} parses the program's syntax tree from the source code to acquire knowledge of the syntactic structure. 
TransCoder-IR \cite{szafranieccode} parses its low-level representation, LLVM IR, from the source code and designs corresponding pre-training tasks to improve code representation. 
As LLMs demonstrate exceptional capabilities, their potential in code translation has also been investigated. 
TRANSAGENT \cite{yuan2024transagent} utilizes LLMs as agents to rectify syntax and semantic errors in code translation. 
LASSI \cite{dearing2024lassi} proposes a self-enhancement method, providing feedback during compilation to the LLM through guided debugging and refactoring prompts. 
Inspired by these works, \Coder utilizes code analysis tools to acquire knowledge related to code executability and injects this knowledge into LLMs to enhance their code translation capability. 

\subsection{Large Language Models for Code}
 With the development of LLMs, language models tailored for code have garnered significant attention in the community. 
 Representative LLMs include CodeLlama \cite{roziere2023code}, Qwen-Coder \cite{hui2024qwen2}, StarCoder \cite{li2023starcoder}, CodeT5+ \cite{wang2023codet5+}, and Deepseek-Coder \cite{guo2024deepseek}. 
 These LLMs benefit from extensive pre-training on large code corpora, resulting in a strong understanding of the semantics of source code. 
 Nevertheless, compared to state-of-the-art closed-source LLMs such as GPT-4 \cite{achiam2023gpt}, these models often lag in capability. 
 To further enhance the ability of code LLMs to address specific coding issues, other works have focused on fine-tuning of pre-trained LLMs. 
 Recent efforts include InstructCoder \cite{li2024instructcoder}, WizardCoder \cite{luo2023wizardcoder}, PanGu-Coder2 \cite{shen2023pangu}, Magicoder \cite{wei2024magicoder}, WaveCoder \cite{yu2024wavecoder},. 
 These LLMs are built upon pre-trained code LLMs and undergo post-training to further enhance model capabilities. 
 However, these code LLMs are not specifically tailored for code translation, they merely learn from source code while neglecting the requirements for executability inherent in code translation. 
 Unlike these works, \Coder encodes executability representation of source code and then employs instruction-tuning to compel LLMs to learn these executability representations, improving their code translation abilities.

\section{Methodology}


\lstset{
  escapeinside={(*@}{@*)}  
}

\lstdefinestyle{lfonts}{
  backgroundcolor=\color{white!5},
  basicstyle   = \footnotesize\ttfamily,
  stringstyle  = \color{purple},
  keywordstyle = \color{green!60!black}\bfseries,
  commentstyle = \color{olive},
}
\lstdefinestyle{lnumbers}{
}
\lstdefinestyle{llayout}{
  breaklines       = true,
  tabsize          = 2,
  columns          = flexible,
}
\lstdefinestyle{lgeometry}{
}
\lstdefinestyle{lgeneral}{
  style = lfonts,
}
\lstdefinestyle{python}{
  language = {Python},
  style    = lgeneral,
}

\begin{figure*}[t]
\centering
\begin{adjustbox}{width=0.95\linewidth,center}
\begin{tabular}{l l l}

\midrule
\footnotesize{Input Code Snippet} & \footnotesize{w/o Functional-semantic Representation} & \footnotesize{with Functional-semantic Representation}\\
\midrule


\begin{lstlisting}[style=Python, xleftmargin=0cm, basicstyle=\fontencoding{T1}\fontfamily{lmtt}\scriptsize]
# C++
int smallest ( int x, int y, int z ) {
  if ( ! ( y / x ) ) 
    return ( ! ( y / z ) ) ? y : z;
  return ( ! ( x / z ) ) ? x : z;
}
\end{lstlisting} & 

\begin{lstlisting}[style=Python, xleftmargin=0cm, basicstyle=\fontencoding{T1}\fontfamily{lmtt}\scriptsize]
# Python
def smallest(x, y, z):
    if not (y % x):
        return y if not (y % z) else z
    return x if not (x % z) else z
\end{lstlisting} & 

\begin{lstlisting}[style=Python, xleftmargin=0cm, basicstyle=\fontencoding{T1}\fontfamily{lmtt}\scriptsize]
# Python
def smallest ( x , y , z ) :
    if ( not ( y / x ) ) :
        return y if not ( y / z ) else z
    return x if not ( x / z ) else z
\end{lstlisting} \\


\midrule
\footnotesize{Input Code Snippet} & \footnotesize{w/o Syntactic-structure Representation} & \footnotesize{with Syntactic-structure Representation}\\
\midrule
\begin{lstlisting}[style=Python, xleftmargin=0cm, basicstyle=\fontencoding{T1}\fontfamily{lmtt}\scriptsize]
# C++
for (int i = num.size() - 1; i >= 0; i--) {
    int digit = num[i] - '0';
    result += digit * series[series_index];
    series_index = (series_index + 1) % 6;
    result %= 7;
}
\end{lstlisting} & 

\begin{lstlisting}[style=Python, xleftmargin=0cm, basicstyle=\fontencoding{T1}\fontfamily{lmtt}\scriptsize]
# Python
for i in range(len(num) - 1, -1, -1):
    digit = int(num[i])
    result += digit * series[series_index]
    series_index = (series_index + 1) % 6
    result %= 7
\end{lstlisting} & 

\begin{lstlisting}[style=Python, xleftmargin=0cm, basicstyle=\fontencoding{T1}\fontfamily{lmtt}\scriptsize]
# Python
for i in range ( len ( num ) - 1 , - 1 , - 1 ) :
    digit = ord ( num [ i ] ) - ord ( '0' )
    result += digit * series [ series_index ]
    series_index = ( series_index + 1 ) % 6
    result %= 7
\end{lstlisting} \\


\midrule
\footnotesize{Input Code Snippet} & \footnotesize{w/o Variable-dependency Representation} & \footnotesize{with Variable-dependency Representation}\\
\midrule
\begin{lstlisting}[style=Python, xleftmargin=0cm, basicstyle=\fontencoding{T1}\fontfamily{lmtt}\scriptsize]
# C++
string minLexRotation ( string st ) {
  int n = st . length ( );
  string arr [ n ];
  string concat = st + st;
  for ( int i = 0; i < n; i ++ ) 
      arr[i] = concat.substr(i, n);
  sort ( arr, arr + n );
  return arr [ 0 ];
}
\end{lstlisting} & 

\begin{lstlisting}[style=Python, xleftmargin=0cm, basicstyle=\fontencoding{T1}\fontfamily{lmtt}\scriptsize]
# Python
def minLexRotation ( st ):
    n = len ( st )
    arr = [ st [ i : ] for i in range ( n ) ]
    arr.sort ( )
    return arr [ 0 ]
\end{lstlisting} & 

\begin{lstlisting}[style=Python, xleftmargin=0cm, basicstyle=\fontencoding{T1}\fontfamily{lmtt}\scriptsize]
# Python
def minLexRotation ( st ) :
    n = len ( st )
    arr = [ None ] * n
    concat = st + st
    for i in range ( n ) :
        arr [ i ] = concat [ i : i + n ]
    arr.sort ( )
    return arr [ 0 ]
\end{lstlisting} \\

    

 \end{tabular}
 \end{adjustbox}
    \caption{\textbf{Three types of Executability Representations.} The first example illustrates an error related to functional semantic, where the baseline model is not informed of the function's role in obtaining the minimum number, substituting a similar modulus symbol for the division operator. The second example highlights an error in syntactic structure, where the baseline model uses forced type conversion to convert characters to numbers, which raises a ValueError exception when non-numeric characters are included in the input. The third example presents an error regarding variable dependency, where the baseline model has not learned the transmission of variables, thereby neglecting to create a concatenation of string with itself. These errors result in a minimal edit distance but have a significant impact on the execution. Learning the executability representation can indicate the execution status, aiding in resolving these issues.}
    
    \vspace{-0.5cm}
    \label{fig:exe_repr_case}
\end{figure*}

\subsection{Overview}
In this paper, we introduce \Coder, a LLM tailored for code translation task, which seeks to improve the capabilities of LLMs in code translation by leveraging executability representations. 
Executability representation refers to information indicating the execution state of the source code, which relies on code analysis tools designed by programming language experts and cannot be derived directly from the source code. 
The automatic translation between different programming languages necessitates that the translated code executes correctly. 
However, current LLMs are pre-trained only on source code, neglecting executability representations closely associated with execution states, leading to inadequate cross-programming language understanding and unreliable automated code translation. 
To address this problem, \Coder first encodes the executability representations of the source code, and then employs instruction fine-tuning to compel LLMs to learn these executability representations, thereby enhancing their cross-programming language understanding and achieving more reliable automated code translation. Figure \ref{fig:overview} illustrates the pipeline of \Coder.


\subsection{Executability Representation for Code Translation}
In order to achieve more reliable code translation, the \Coder first customizes an executability knowledge representation strategy specifically for LLMs, encoding three types of executability representations, including functional semantic, syntactic structure, and variable dependency. 
Then, the \Coder constructs a specialized instruction fine-tuning dataset, XLCoST-Instruct, based on the cross-programming language alignment dataset XLCoST \cite{zhu2022xlcost}. 
XLCoST-Instruct compels LLMs to learn the executability representations of the source code, enhancing the cross-language understanding of LLMs.

\textbf{Functional-semantic Representation.} In different programming languages, code with the same functionality may exhibit significant differences in form. Therefore, learning solely from the semantics of the source code may lead to misunderstandings of the code's functionality, which may result in erroneous code translation and execution failures, as shown in Figure \ref{fig:exe_repr_case}. To address this, the \Coder encodes the functional-semantic representation of the source code, in order to align the functionalities of the source and target code. Functional semantic refers to the effect of the source code once it has been executed, and aligning the code functionalities of the source and target languages guarantees that their execution results remain consistent.

\Coder encodes the functional semantics of source code using a natural language description of its functionality, as shown in Figure \ref{fig:tuning_nl}. 
LLMs are pre-trained on a large amount of natural language data, which enables them to have a good understanding of natural language, facilitating their learning of functional semantics \cite{zhao2023survey}. 
In fact, natural language descriptions that articulate the functional semantics of code are readily available. 
A substantial amount of public code data resources is collected from open-source repositories (such as GitHub), where comments accompanying commits serve as high-quality functional semantics of code \cite{ahmad2021unified}.

\textbf{Syntactic-structure Representation.} Compared to natural languages, programming languages have clear syntactic structures and strict grammatical rules. Merely learning the context of source code while neglecting its syntactic structure may result in translated code using incorrect syntax, leading to unexpected execution results or compilation errors, as shown in Figure \ref{fig:exe_repr_case}. To address this issue, \Coder encodes the syntactic structure information of source code using Abstract Syntax Tree (AST) to enhance LLMs' understanding of the syntax of different programming languages. An AST is a tree that represents the abstract syntactic structure of source code, where each subtree represents a continuous range of subword tokens, and each leaf node represents a single token; this structure has been shown to effectively encode the syntactic structure of source code \cite{gongast}. To construct the AST, \Coder utilizes the lightweight multilingual parser tree-sitter \cite{tree-sitter} to parse the source code.

However, AST is structured graph data, and LLMs are pre-trained only on unstructured text, resulting in limited understanding of graph structure \cite{tang2024graphgpt}, making it difficult to learn the syntactic structure of source code. To address this, inspired by \cite{fatemitalk}, \Coder further encodes the AST into unstructured text that is easier for LLMs to understand, serving as a representation of the syntactic structure, as shown in Figure \ref{fig:tuning_ast}. 
\Coder simplifies ASTs by keeping only leaf node tokens. It then indexes the AST as a graph and describes nodes and edges in natural language, enhancing LLM understanding of code structure \cite{fatemitalk}. Further details are provided in Appendix \ref{sec:ast_case}. 

\textbf{Variable-dependency Representation.} In different programming languages, the same variable often has different semantics due to variations in programmer preferences and naming conventions \cite{cheng2024dataflow}. The diversity of variable semantics may lead to incorrect variable dependency relationships when solely learning the semantics of source code, further resulting in anomalous or unexpected execution outcomes, as shown in Figure \ref{fig:exe_repr_case}. To address this issue, \Coder encodes the variable dependency information of the source code using Data Flow Graph (DFG) to enhance LLMs' understanding of variable dependencies in code. A DFG is a graph that describes the dependencies and interactions between variables, where each node represents a variable and each edge indicates the source of these variables. To construct the DFG, \Coder utilizes the lightweight multilingual parser tree-sitter \cite{tree-sitter} to parse the source code.

Given that a DFG is structured graph data, similar to the handling of AST, \Coder also encodes the DFG into unstructured text that is easier for LLMs to understand, serving as a representation of variable dependencies, as shown in Figure \ref{fig:tuning_dfg}. \Coder directly assigns a numerical index to each node in the DFG and then uses natural language to represent the node information and edge information of the graph. For the node information, \Coder describes the token content of each node. For the edge information, \Coder describes the neighboring nodes of each node.

\subsection{XLCoST-Instruct}
To enable LLMs to learn the executability representation and enhance their cross-language understanding capabilities, \Coder developed a specially designed instruction fine-tuning dataset XLCoST-Instruct for code translation. Details of XLCoST-Instruct are in Appendix \ref{sec:tuning_dataset} and \ref{sec:tuning_case}.

\subsection{Progressive Executability Representation Learning}

\Coder aims to utilize executability representations such as functional semantics, syntactic structures, and variable dependencies in code to enhance the capabilities of LLMs in code translation tasks. However, the acquisition of executability representations relies on code analysis tools, which, unlike source code, are domain-specific high-level knowledge that is difficult to comprehend directly, facing two significant challenges in the learning process. First, how can we learn both low-level code semantics and high-level executability representations simultaneously? Second, how can we design learning strategies to enhance the understanding of high-level executability representations? To address these issues, \Coder proposes a Progressive Executability Representation Learning (PERL) to leverage progressively refined executability representations of code to enhance LLMs' cross-language understanding capabilities.

Specifically, to address the first issue, PERL incorporates executability representations as auxiliary knowledge during instruction fine-tuning, as illustrated in Appendix \ref{sec:tuning_case}. PERL concatenates the source code with its corresponding executability representation and then prompts LLMs to translate code based on the respective executability representations. This strategy has been shown to effectively enable LLMs to learn high-level knowledge \cite{yin2023natural}.

To address the second issue, PERL has designed a phased fine-tuning strategy. Research \cite{robins2003learning} indicates that the optimal learning of programming skills is progressive, beginning with an understanding of the high-level intentions of a program, followed by the implementation of low-level code structures. Inspired by this, PERL has designed a phased fine-tuning strategy where each fine-tuning stage independently learns different executability representations. The key idea is that the functional semantics, syntactic structures, and variable dependencies of source code are executability representations that are refined progressively. Phased progressive learning aligns with the learning theories of programming experts, facilitating representational learning. Specifically, \Coder divides the instruction fine-tuning process into four stages: source code, functional semantics, syntactic structures, and variable dependencies, with each stage fine-tuning using only one type of data. Once the fine-tuning process of one stage converges, the next stage of fine-tuning is initiated.
\section{Experimental Evaluation}

\begin{table*}[htbp]
  \centering
  \scalebox{0.6}{
    \begin{tabular}{cccccccccc}
    \hline\hline
    \multicolumn{10}{c}{\textbf{Compilation Accuracy (CA)}} \\
    \midrule
    \textbf{Model} & Deepseek-Coder & CodeLLama & Magicoder & Qwen2.5-Coder & WaveCoder & GPT-3.5 & GPT-4 & \multicolumn{1}{c|}{GPT-4o} & Ours \\
    \hline
    \textbf{From C++} & 88.143 & 58.836 & 80.481 & 66.106 & 72.586 & 85.330 & \textbf{92.698} & \multicolumn{1}{c|}{91.458} & 91.559 \\
    \textbf{To C++} & 91.328 & 37.901 & 87.259 & 54.711 & 92.077 & 86.724 & 94.218 & \multicolumn{1}{c|}{90.792} & \textbf{97.002} \\
    \textbf{From Python} & 84.548 & 29.810 & 82.407 & 46.344 & 86.235 & 86.342 & 87.702 & \multicolumn{1}{c|}{88.018} & \textbf{90.995} \\
    \textbf{To Python} & 70.582 & 81.466 & 54.957 & 53.987 & 49.138 & 84.591 & 89.763 & \multicolumn{1}{c|}{91.272} & \textbf{93.534} \\
    \textbf{From Java} & 72.760 & 70.930 & 63.182 & 48.956 & 68.019 & 84.433 & 91.498 & \multicolumn{1}{c|}{89.150} & \textbf{95.482} \\
    \textbf{To Java} & 83.542 & 40.208 & 83.854 & 52.708 & 85.625 & 84.792 & \textbf{87.917} & \multicolumn{1}{c|}{86.563} & 87.500 \\
    \textbf{Average} & 81.817 & 53.192 & 75.357 & 53.802 & 75.613 & 85.369 & 90.633 & \multicolumn{1}{c|}{89.542} & \textbf{92.679} \\
    \hline\hline
    \multicolumn{10}{c}{\textbf{Case Computational Accuracy (CCA)}} \\
    \midrule
    \textbf{Model} & Deepseek-Coder & CodeLLama & Magicoder & Qwen2.5-Coder & WaveCoder & GPT-3.5 & GPT-4 & \multicolumn{1}{c|}{GPT-4o} & Ours \\
    \hline
    \textbf{From C++} & 83.119 & 53.488 & 76.133 & 63.335 & 68.346 & 80.633 & \textbf{88.141} & \multicolumn{1}{c|}{86.912} & 87.221 \\
    \textbf{To C++} & 85.535 & 35.557 & 81.488 & 52.045 & 86.456 & 81.724 & 90.064 & \multicolumn{1}{c|}{86.403} & \textbf{91.970} \\
    \textbf{From Python} & 79.340 & 28.059 & 76.641 & 44.251 & 80.348 & 81.509 & 83.540 & \multicolumn{1}{c|}{83.340} & \textbf{85.891} \\
    \textbf{To Python} & 64.250 & 71.510 & 49.932 & 50.586 & 43.847 & 77.844 & 82.723 & \multicolumn{1}{c|}{84.391} & \textbf{86.590} \\
    \textbf{From Java} & 67.721 & 64.135 & 58.896 & 46.513 & 63.849 & 79.051 & 86.513 & \multicolumn{1}{c|}{83.937} & \textbf{89.948} \\
    \textbf{To Java} & 80.396 & 38.615 & 80.250 & 51.469 & 82.240 & 81.625 & \textbf{85.406} & \multicolumn{1}{c|}{83.396} & 84.500 \\
    \textbf{Average} & 76.727 & 48.560 & 70.557 & 51.367 & 70.848 & 80.397 & 86.064 & \multicolumn{1}{c|}{84.730} & \textbf{87.687} \\
    \hline\hline
    \multicolumn{10}{c}{\textbf{Test Computational Accuracy (TCA)}} \\
    \hline
    \textbf{Model} & Deepseek-Coder & CodeLLama & Magicoder & Qwen2.5-Coder & WaveCoder & GPT-3.5 & GPT-4 & \multicolumn{1}{c|}{GPT-4o} & Ours \\
    \hline
    \textbf{From C++} & 78.434 & 50.467 & 71.861 & 60.873 & 64.494 & 77.126 & \textbf{84.145} & \multicolumn{1}{c|}{82.813} & 83.226 \\
    \textbf{To C++} & 80.086 & 32.548 & 76.017 & 49.465 & 80.835 & 76.981 & 84.904 & \multicolumn{1}{c|}{80.942} & \textbf{86.617} \\
    \textbf{From Python} & 74.580 & 26.326 & 71.495 & 41.986 & 74.886 & 76.695 & 78.275 & \multicolumn{1}{c|}{77.537} & \textbf{80.717} \\
    \textbf{To Python} & 59.267 & 65.948 & 46.013 & 47.414 & 40.086 & 73.276 & 77.478 & \multicolumn{1}{c|}{79.203} & \textbf{80.927} \\
    \textbf{From Java} & 63.631 & 59.100 & 55.237 & 44.228 & 60.500 & 75.081 & 82.254 & \multicolumn{1}{c|}{79.691} & \textbf{85.163} \\
    \textbf{To Java} & 77.292 & 37.396 & 76.563 & 50.208 & 78.958 & 78.646 & \textbf{82.292} & \multicolumn{1}{c|}{79.896} & 81.563 \\
    \textbf{Average} & 72.215 & 45.297 & 66.198 & 49.029 & 66.627 & 76.301 & 81.558 & \multicolumn{1}{c|}{80.014} & \textbf{83.035} \\
    \hline\hline
    \end{tabular}%
    }
    \caption{\textbf{Execution-based evaluation} of ExeCoder and baseline models on TransCoder-test-X.}
  \label{tab:baseline_exe}%
  \vspace{-0.5cm}
\end{table*}%

\subsection{Experimental Setup}
\textbf{Dataset.} We conducted comprehensive experiments on the widely used code translation public dataset TransCoder-test \cite{transcoder} to demonstrate the effectiveness of the \Coder. TransCoder-test contains solutions to programming problems collected from GeeksForGeeks, with each problem implemented in C++, Java, and Python, totaling 948 parallel samples. TransCoder-test is equipped with predefined unit test templates for over half of the translation pairs to evaluate whether the generated functions return the same output as the reference functions given the same input. Each unit test consists of ten test cases.

However, these unit test templates establish fixed parameter passing methods or return types that can only evaluate specific implementations. When the generated functions do not conform to the predefined implementations, the unit tests cannot execute as expected, even if the generated functions are functionally equivalent to the reference functions, resulting in unreliable evaluation. To address this, we enhanced TransCoder-test to ensure that the unit test results are capable of assessing the code translation ability. 
The enhanced test set is referred to as TransCoder-test-X, with detailed processing cases provided in Appendix \ref{sec:TransCoder-test-X}.


\textbf{Match-based Metrics.} The matching-based metrics aim to evaluate the quality of generated code through static code analysis. Following existing work \cite{huang2023program}, we employed three metrics: Exact Match (EM), BLEU \cite{papineni2002bleu}, and CodeBLEU \cite{ren2020codebleu} for evaluation. These metrics can measure the n-gram overlap, syntactic structure, and semantic equivalence of the generated code.

\textbf{Execution-based Metrics.} The execution-based metrics aim to assess the quality of generated code through the execution status of the code. Following existing work \cite{transcoder}, we utilize test computational accuracy~(TCA) as a metric, which measures the proportion of code that passes unit tests. Furthermore, to further reflect the fine-grained execution status of the code, we have designed two additional metrics: Compilation Accuracy~(CA) and Case Computational Accuracy~(CCA), the former measuring the proportion of code that successfully compiles, and the latter measuring the proportion of code that passes case tests.

\textbf{Baselines.} To assess the effectiveness of \Coder, we compare it with state-of-the-art LLMs. Specifically, we use three advanced closed-source models as baselines: OpenAI's GPT-3.5, GPT-4, and GPT-4-o \cite{achiam2023gpt}. Additionally, we select five leading open-source code LLMs as baselines: Deepseek-Coder-6.7b-instruct (our base model) \cite{guo2024deepseek}, CodeLLama-7B \cite{roziere2023code}, Magicoder-S-DS-6.7B \cite{wei2024magicoder}, Qwen2.5-Coder-7B \cite{hui2024qwen2}, and Wavecoder-ultra-6.7b \cite{yu2024wavecoder}.



\subsection{Experimental Results}
\textbf{Effectiveness of \Coder.} To assess the effectiveness of \Coder in code translation task, we evaluated \Coder on the TransCoder-test-X, which includes translations among the three most commonly used programming languages: C++, Python, and Java. Tables \ref{tab:baseline_exe} and \ref{tab:baseline_match} present the comparison results with baseline models. 

Generally speaking, \Coder achieved state-of-the-art performance in code translation task. Compared to the state-of-the-art closed-source model GPT-4, \Coder outperformed it by 2.05\%, 1.62\%, and 1.48\% in execution-based metrics, and by 3.78\%, 58.72\%, and 29.24\% in match-based metrics. This evidence demonstrates \Coder's strong effectiveness in code translation task. Compared to open-source models, \Coder achieved even more significant advantages, outperforming by up to 39.49\%, 39.13\%, and 37.74\% in execution-based metrics, and by 4.38\%, 68.55\%, and 61.78\% in match-based metrics. This significant leap indicates that \Coder has reached state-of-the-art performance in code translation task. Furthermore, compared to the foundation model Deepseek-Coder-6.7b-instruct, \Coder also attained significant advantages, outperforming by 10.86\%, 10.96\%, and 10.82\% in execution-based metrics, and by 3.78\%, 58.72\%, and 29.24\% in match-based metrics. These advantages highlight that the executability representation facilitates enhanced cross-programming language understanding of LLMs, achieving precise and reliable automated code translation.

\begin{figure}[htbp]
\centerline{
\includegraphics[width=\linewidth]{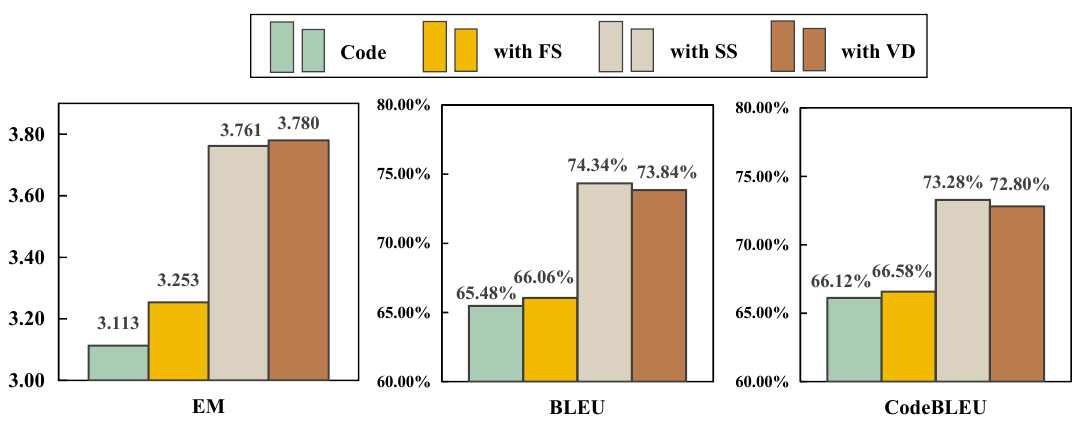}
}
\caption{Match-based evaluation of different variants of ExeCoder on TransCoder-test-X.}
\label{repr_ablation}
\vspace{-0.4cm}
\end{figure}

\textbf{Effectiveness of Executability Representation.} We conduct an ablation study to evaluate the effectiveness of the executability representation. Specifically, we examine the following variants of \Coder, where the fine-tuning process for each variant maintains the same step. 
\begin{itemize}
\item \textbf{Code}: Fine-tuning using only the source code.
\item \textbf{with FS}: Fine-tuning using only the source code and functional semantic representation.
\item \textbf{with SS}: Fine-tuning using only the source code and syntactic structure representation.
\item \textbf{with VD}: Fine-tuning using only the source code and variable dependency representation.
\end{itemize}

Tables \ref{tab:repr_ca}, \ref{tab:repr_cca}, \ref{tab:repr_tca} and Figure \ref{repr_ablation} present our results, which indicate that the learning from each executability representation contributes to varying degrees of improvement. For execution-based metrics, the inclusion of the three different types of executability representations resulted in an average increase of 0.41\%, 1.64\%, and 1.21\%. For match-based metrics, the integration of the three different types of executability representations led to an average increase of 0.39\%, 5.56\%, and 5.24\%. These advantages indicate that executability representations can provide execution status beyond code semantics, enhancing the cross-language understanding capability of LLMs.

\begin{table}[!htbp]
  \centering
    \resizebox{\linewidth}{!}{
    \begin{tabular}{ccccc}
    \midrule
    \textbf{Model} & Code  & with FS & with SS & with VD \\
    \midrule
    \textbf{From C++} &  91.13 &  91.56 (+0.43) &  91.66 (+0.53) &  91.14 (+0.01)  \\
    \textbf{To C++} &  96.36 &  96.47 (+0.11) &  96.90 (+0.54) &  97.11 (+0.75)  \\
    \textbf{From Python} &  89.21 &  89.52 (+0.31) &  91.20 (+1.99) &  90.78 (+1.57)  \\
    \textbf{To Python} &  90.41 &  91.59 (+1.18) &  93.00 (+2.59) &  92.89 (+2.48)  \\
    \textbf{From Java} &  92.68 &  93.54 (+0.86) &  95.05 (+2.37) &  95.27 (+2.59)  \\
    \textbf{To Java} &  86.25 &  86.56 (+0.31) &  88.02 (+1.77) &  87.19 (+0.94)  \\
    \midrule
    \textbf{Average} &  91.01 &  91.54 (+0.53) &  92.64 (+1.63) &  92.39 (+1.38)  \\
    \midrule
    \end{tabular}%
    }
     \caption{\textbf{Compilation Accuracy (CA)} of different variants of ExeCoder.}
  \label{tab:repr_ca}%
  \vspace{-0.5cm}
\end{table}%
\begin{table}[!htbp]
  \centering
  \resizebox{\linewidth}{!}{
    \begin{tabular}{ccccc}
    \midrule
    \textbf{Model} & Code  & with FS & with SS & with VD \\
    \midrule
    \textbf{From C++} &  86.04 &  86.48 (+0.44) &  87.33 (+1.29) &  86.56 (+0.52)  \\
    \textbf{To C++} &  91.21 &  90.89 (-0.32) &  91.68 (+0.47) &  91.78 (+0.57)  \\
    \textbf{From Python} &  84.27 &  84.28 (+0.01) &  85.37 (+1.10) &  84.79 (+0.52)  \\
    \textbf{To Python} &  83.11 &  83.96 (+0.85) &  85.83 (+2.72) &  85.59 (+2.48)  \\
    \textbf{From Java} &  87.13 &  87.43 (+0.30) &  89.38 (+2.25) &  89.38 (+2.25)  \\
    \textbf{To Java} &  83.11 &  83.33 (+0.22) &  84.57 (+1.46) &  83.36 (+0.25)  \\
    \midrule
    \textbf{Average} &  85.81 &  86.06 (+0.25) &  87.36 (+1.55) &  86.91 (+1.10)  \\
    \midrule
    \end{tabular}%
    }
     \caption{\textbf{Case Computational Accuracy (CCA)} of different variants of ExeCoder.}
  \label{tab:repr_cca}%
  \vspace{-0.5cm}
\end{table}%
\begin{table}[H]
  \centering
  \resizebox{\linewidth}{!}{
    \begin{tabular}{ccccc}
    \midrule
    \textbf{Model} & Code  & with FS & with SS & with VD \\
    \midrule
    \textbf{From C++} &  81.62 &  82.17 (+0.55) &  82.90 (+1.28) &  82.16 (+0.54)  \\
    \textbf{To C++} &  85.22 &  85.12 (-0.10) &  86.19 (+0.97) &  86.08 (+0.86)  \\
    \textbf{From Python} &  78.28 &  78.60 (+0.32) &  80.08 (+1.80) &  78.93 (+0.65)  \\
    \textbf{To Python} &  76.40 &  77.59 (+1.19) &  79.31 (+2.91) &  79.20 (+2.80)  \\
    \textbf{From Java} &  81.82 &  82.25 (+0.43) &  83.98 (+2.16) &  84.09 (+2.27)  \\
    \textbf{To Java} &  80.10 &  80.31 (+0.21) &  81.46 (+1.36) &  79.90 (-0.20)  \\
    \midrule
    \textbf{Average} &  80.58 &  81.01 (+0.43) &  82.32 (+1.74) &  81.73 (+1.15)  \\
    \midrule
    \end{tabular}%
    }
    \caption{\textbf{Test Computational Accuracy (TCA)} of different variants of ExeCoder.}
  \label{tab:repr_tca}%
  \vspace{-0.5cm}
\end{table}%

\textbf{Effectiveness of Progressive Executability Representation Learning.} We conduct an ablation study to evaluate the effectiveness of the PERL strategy. Specifically, we assess the impact of each fine-tuning phase in PERL on the model's code translation capability. 
Tables \ref{tab:perl_ca}, \ref{tab:perl_cca}, and \ref{tab:perl_tca} present our results. 
We also conducted a more detailed ablation study on PERL, with details provided in Appendix \ref{sec:perl} due to space limitations.

Overall, \Coder's meticulously designed PERL strategy significantly enhances the code translation performance through the learning of executability representations. Compared to models fine-tuned solely on source code, the inclusion of executability representations raises the upper limits of model capability, with three execution-based metrics improving by 0.53\%, 0.85\%, and 1.28\%, respectively. Furthermore, the learning of specific executability representations at each stage improves the model's cross-language understanding. Relative to the previous stage, each fine-tuning stage further improved the key execution-based metric TCA by 0.50\%, 0.30\%, and 0.49\%, respectively. 

\begin{table}[htbp]
  \centering
  \resizebox{\linewidth}{!}{
    \begin{tabular}{ccccc}
    \midrule
    \textbf{Model} & Code  & Code-FS & Code-FS-SS & Code-FS-SS-VD \\
    \midrule
    \textbf{From C++} &  91.55 &  91.13 (-0.42) &  91.15 (+0.02) &  91.56 (+0.41)  \\
    \textbf{To C++} &  96.36 &  96.36 (+0.00) &  97.11 (+0.75) &  97.00 (-0.11)  \\
    \textbf{From Python} &  89.73 &  90.78 (+1.05) &  90.58 (-0.20) &  91.00 (+0.42)  \\
    \textbf{To Python} &  93.00 &  92.56 (-0.44) &  93.53 (+0.97) &  93.53  (+0.00) \\
    \textbf{From Java} &  95.16 &  94.62 (-0.54) &  95.37 (+0.75) &  95.48 (+0.11)  \\
    \textbf{To Java} &  87.08 &  87.60 (+0.52) &  86.46 (-1.14) &  87.50 (+1.04)  \\
    \midrule
    \textbf{Average} &  92.15 &  92.18 (+0.03) &  92.37 (+0.19) &  92.68 (+0.31)  \\
    \midrule
    \end{tabular}%
    }
    \caption{\textbf{Compilation Accuracy (CA)} of various stages within PERL.}
  \label{tab:perl_ca}%
  \vspace{-0.5cm}
\end{table}%
\begin{table}[htbp]
  \centering

  \resizebox{\linewidth}{!}{
    \begin{tabular}{ccccc}
    \midrule
    \textbf{Model} & Code  & Code-FS & Code-FS-SS & Code-FS-SS-VD \\
    \midrule
    \textbf{From C++} &  86.52 &  86.31 (-0.21) &  86.63 (+0.32) &  87.22 (+0.59)  \\
    \textbf{To C++} &  90.97 &  91.08 (+0.11) &  92.09 (+1.01) &  91.97 (-0.12)  \\
    \textbf{From Python} &  84.65 &  85.77 (+1.12) &  85.38 (-0.38) &  85.89 (+0.51)  \\
    \textbf{To Python} &  85.62 &  85.24 (-0.38) &  86.11 (+0.87) &  86.59 (+0.48)  \\
    \textbf{From Java} &  89.33 &  88.42 (-0.91) &  89.71 (+1.29) &  89.95 (+0.24)  \\
    \textbf{To Java} &  83.91 &  84.48 (+0.57) &  83.53 (-0.95) &  84.50 (+0.97)  \\
    \midrule
    \textbf{Average} &  86.83 &  86.88 (+0.05) &  87.24 (+0.36) &  87.69 (+0.45)  \\
    \midrule
    \end{tabular}%
    }
    \caption{\textbf{Case Computational Accuracy (CCA)} of various stages within PERL.}
  \label{tab:perl_cca}%
  \vspace{-0.5cm}
\end{table}%
\begin{table}[H]
  \centering

  \resizebox{\linewidth}{!}{
    \begin{tabular}{ccccc}
    \midrule
    \textbf{Model} & Code  & Code-FS & Code-FS-SS & Code-FS-SS-VD \\
    \midrule
    \textbf{From C++} &  82.37 &  82.37 (+0.00) &  82.38 (+0.01) &  83.23 (+0.84)  \\
    \textbf{To C++} &  85.01 &  85.87 (+0.86) &  87.05 (+1.18) &  86.62 (-0.43)  \\
    \textbf{From Python} &  78.80 &  80.50 (+1.70) &  80.31 (-0.19) &  80.72 (+0.41)  \\
    \textbf{To Python} &  79.31 &  79.42 (+0.11) &  80.17 (+0.75) &  80.93 (+0.76)  \\
    \textbf{From Java} &  84.09 &  83.87 (-0.21) &  84.95 (+1.07) &  85.16 (+0.22)  \\
    \textbf{To Java} &  80.94 &  81.46 (+0.52) &  80.42 (-1.04) &  81.56 (+1.15)  \\
    \midrule
    \textbf{Average} &  81.75 &  82.25 (+0.50) &  82.55 (+0.30) &  83.03 (+0.49)  \\
    \midrule
    \end{tabular}%
    }
\caption{\textbf{Test Computational Accuracy (TCA)} of various stages within PERL.}
  \label{tab:perl_tca}%
  \vspace{-0.5cm}
\end{table}%

\section{Conclusion}
In this paper, we propose \Coder, a large language model specifically designed for code translation. The key idea of \Coder is to enhance the capabilities of LLMs in code translation by leveraging executability representations such as functional semantics, syntactic structure, and variable dependencies in code. Additionally, we have enhanced the widely used code translation benchmark TransCoder-test, resulting in a new benchmark called TransCoder-test-X, which is capable of evaluating the code translation abilities of LLMs. Evaluation on TransCoder-test-X indicates that \Coder achieves state-of-the-art performance in code translation, surpassing existing open-source large code models by more than 10.88\% to 38.78\% and more than 27.44\% to 42.97\% on two different metrics. Notably, the \Coder outperforms renowned closed-source LLMs, including GPT-3.5, GPT-4, and GPT-4o, highlighting the significant role of executability representations in code translation. In the future, we plan to incorporate richer executability representations for LLMs using code analysis tools.

\section*{Limitations}

A primary limitation of this work is identified. ExeCoder employs static analysis tools to derive executability representations from source code, and these representations often possess extensive contextual information. Although the number of tokens during the fine-tuning process stays within the model's context window, it nonetheless elevates the relevant fine-tuning expenditures, consequently diminishing ExeCoder's scalability to a certain extent. Addressing the reduction of these costs falls outside the purview of the current study and is not the central topic of this paper.



\bibliography{custom}

\newpage
\appendix
\onecolumn

\section{Ablation of PERL}
\label{sec:perl}
we conducted further ablation studies on PERL to demonstrate its necessity and effectiveness. Specifically, we adopted the single-stage mixed training approach and reversed the order of the various representations in PERL, with the evaluation results presented in the table \ref{tab:perl_ablation}. The results show that both training with mixed representations and altering the order of the various representations had varying degrees of impact on the performance of ExeCoder. The training method PERL, which aligns with programming learning theory, achieved optimal performance.

\begin{table}[htbp]
  \centering
    \scalebox{0.7}{
    \begin{tabular}{cccccccr}
    \hline\hline
    \multicolumn{8}{c}{\textbf{Compilation Accuracy (CA)}} \\
    \midrule
    \textbf{Model} & \textbf{NL-AST-DFG (ExeCoder)} & \textbf{NL-DFG-AST} & \textbf{AST-NL-DFG} & \textbf{AST-DFG-NL} & \textbf{DFG-AST-NL} & \textbf{DFG-NL-AST} & \multicolumn{1}{c}{\textbf{Shuffle}} \\
    \midrule
    \textbf{From C++} & 91.559 & 91.354 & 91.347 & 91.243 & 91.782 & 91.020 & \multicolumn{1}{c}{91.861} \\
    \textbf{To C++} & 97.002 & 96.681 & 96.681 & 96.574 & 96.467 & 95.824 & \multicolumn{1}{c}{96.253} \\
    \textbf{From Python} & 90.995 & 90.579 & 89.948 & 89.844 & 89.948 & 89.410 & \multicolumn{1}{c}{90.460} \\
    \textbf{To Python} & 93.534 & 93.534 & 93.427 & 93.427 & 93.103 & 92.888 & \multicolumn{1}{c}{93.211} \\
    \textbf{From Java} & 95.482 & 95.053 & 95.375 & 95.268 & 94.299 & 95.053 & \multicolumn{1}{c}{95.268} \\
    \textbf{To Java} & 87.500 & 86.771 & 86.563 & 86.354 & 86.458 & 86.771 & \multicolumn{1}{c}{88.125} \\
    \midrule
    \textbf{Average} & 92.679 & 92.329 & 92.223 & 92.118 & 92.010 & 91.828 & \multicolumn{1}{c}{92.530} \\
    \hline\hline
    \multicolumn{8}{c}{\textbf{Case Computational Accuracy (CCA)}} \\
    \midrule
    \textbf{Model} & \textbf{NL-AST-DFG (ExeCoder)} & \textbf{NL-DFG-AST} & \textbf{AST-NL-DFG} & \textbf{AST-DFG-NL} & \textbf{DFG-AST-NL} & \textbf{DFG-NL-AST} & \multicolumn{1}{c}{\textbf{Shuffle}} \\
    \midrule
    \textbf{From C++} & 87.221 & 87.145 & 86.892 & 86.548 & 86.880 & 86.247 & \multicolumn{1}{c}{87.616} \\
    \textbf{To C++} & 91.970 & 91.842 & 91.713 & 91.842 & 91.028 & 90.664 & \multicolumn{1}{c}{90.867} \\
    \textbf{From Python} & 85.891 & 85.542 & 84.940 & 84.807 & 84.335 & 84.334 & \multicolumn{1}{c}{84.682} \\
    \textbf{To Python} & 86.590 & 86.396 & 86.353 & 85.868 & 85.017 & 85.491 & \multicolumn{1}{c}{85.847} \\
    \textbf{From Java} & 89.948 & 89.186 & 89.745 & 89.615 & 88.184 & 89.251 & \multicolumn{1}{c}{89.218} \\
    \textbf{To Java} & 84.500 & 83.635 & 83.510 & 83.260 & 83.354 & 83.677 & \multicolumn{1}{c}{84.802} \\
    \midrule
    \textbf{Average} & 87.687 & 87.291 & 87.192 & 86.990 & 86.466 & 86.611 & \multicolumn{1}{c}{87.172} \\
    \hline\hline
    \multicolumn{8}{c}{\textbf{Test Computational Accuracy (TCA)}} \\
    \midrule
    \textbf{Model} & \textbf{NL-AST-DFG (ExeCoder)} & \textbf{NL-DFG-AST} & \textbf{AST-NL-DFG} & \textbf{AST-DFG-NL} & \textbf{DFG-AST-NL} & \textbf{DFG-NL-AST} & \multicolumn{1}{c}{\textbf{Shuffle}} \\
    \midrule
    \textbf{From C++} & 83.226 & 82.813 & 83.226 & 82.583 & 82.475 & 82.148 & \multicolumn{1}{c}{83.308} \\
    \textbf{To C++} & 86.617 & 86.938 & 86.617 & 86.831 & 85.760 & 85.760 & \multicolumn{1}{c}{85.439} \\
    \textbf{From Python} & 80.717 & 80.630 & 79.991 & 79.678 & 78.935 & 79.557 & \multicolumn{1}{c}{79.340} \\
    \textbf{To Python} & 80.927 & 80.172 & 80.603 & 79.849 & 78.556 & 79.095 & \multicolumn{1}{c}{79.310} \\
    \textbf{From Java} & 85.163 & 83.980 & 84.733 & 84.732 & 83.011 & 84.088 & \multicolumn{1}{c}{83.872} \\
    \textbf{To Java} & 81.563 & 80.313 & 80.729 & 80.313 & 80.104 & 80.938 & \multicolumn{1}{c}{81.771} \\
    \midrule
    \textbf{Average} & 83.035 & 82.474 & 82.650 & 82.331 & 81.473 & 81.931 & \multicolumn{1}{c}{82.173} \\
    \hline\hline
    \end{tabular}%
    }
    \caption{Additional Ablation of PERL.}
  \label{tab:perl_ablation}%
\end{table}%

\newpage
\section{Match-based evaluation}

\begin{table*}[htbp]
  \centering
  \scalebox{0.75}{
    \begin{tabular}{cccccccccc}
    \hline\hline
    \multicolumn{10}{c}{\textbf{Exact Match (EM)}} \\
    \midrule
    \textbf{Model} & Deepseek-Coder & CodeLLama & Magicoder & Qwen2.5-Coder & WaveCoder & GPT-3.5 & GPT-4 & \multicolumn{1}{c|}{GPT-4o} & Ours \\
    \hline
    \textbf{From C++} & 0.000 & 0.055 & 0.000 & 0.000 & 0.000 & 0.160 & 0.000 & \multicolumn{1}{c|}{0.000} & \textbf{6.120} \\
    \textbf{To C++} & 0.000 & 0.265 & 0.160 & 0.000 & 0.000 & 1.635 & 0.055 & \multicolumn{1}{c|}{0.370} & \textbf{2.325} \\
    \textbf{From Python} & 0.000 & 0.000 & 0.000 & 0.000 & 0.000 & 0.000 & 0.000 & \multicolumn{1}{c|}{0.000} & \textbf{0.055} \\
    \textbf{To Python} & 0.000 & 0.000 & 0.000 & 0.000 & 0.000 & 0.000 & 0.000 & \multicolumn{1}{c|}{0.000} & \textbf{10.445} \\
    \textbf{From Java} & 0.000 & 0.265 & 0.160 & 0.000 & 0.000 & 1.635 & 0.055 & \multicolumn{1}{c|}{0.370} & \textbf{6.965} \\
    \textbf{To Java} & 0.000 & 0.055 & 0.000 & 0.000 & 0.000 & 0.160 & 0.000 & \multicolumn{1}{c|}{0.000} & \textbf{0.370} \\
    \textbf{Average} & 0.000 & 0.107 & 0.053 & 0.000 & 0.000 & 0.598 & 0.018 & \multicolumn{1}{c|}{0.123} & \textbf{4.380} \\
    \hline\hline
    \multicolumn{10}{c}{\textbf{BLEU}} \\
    \midrule
    \textbf{Model} & Deepseek-Coder & CodeLLama & Magicoder & Qwen2.5-Coder & WaveCoder & GPT-3.5 & GPT-4 & \multicolumn{1}{c|}{GPT-4o} & Ours \\
    \hline
    \textbf{From C++} & 9.041 & 16.316 & 9.821 & 4.778 & 8.696 & 9.639 & 7.826 & \multicolumn{1}{c|}{8.846} & \textbf{72.627} \\
    \textbf{To C++} & 11.486 & 10.305 & 10.110 & 3.789 & 7.820 & 10.534 & 8.857 & \multicolumn{1}{c|}{23.432} & \textbf{80.600} \\
    \textbf{From Python} & 7.601 & 8.926 & 7.898 & 4.908 & 7.565 & 7.938 & 7.120 & \multicolumn{1}{c|}{7.628} & \textbf{62.708} \\
    \textbf{To Python} & 9.471 & 10.443 & 9.783 & 3.313 & 9.417 & 10.244 & 9.474 & \multicolumn{1}{c|}{9.989} & \textbf{80.765} \\
    \textbf{From Java} & 12.298 & 11.429 & 10.935 & 1.736 & 8.706 & 11.359 & 9.690 & \multicolumn{1}{c|}{24.439} & \textbf{81.735} \\
    \textbf{To Java} & 7.983 & 15.923 & 8.760 & 4.319 & 7.730 & 8.159 & 6.305 & \multicolumn{1}{c|}{7.493} & \textbf{55.705} \\
    \textbf{Average} & 9.647 & 12.224 & 9.551 & 3.807 & 8.322 & 9.646 & 8.212 & \multicolumn{1}{c|}{13.638} & \textbf{72.356} \\
    \hline\hline
    \multicolumn{10}{c}{\textbf{CodeBLEU}} \\
    \midrule
    \textbf{Model} & Deepseek-Coder & CodeLLama & Magicoder & Qwen2.5-Coder & WaveCoder & GPT-3.5 & GPT-4 & \multicolumn{1}{c|}{GPT-4o} & Ours \\
    \hline
    \textbf{From C++} & 37.580 & 36.493 & 9.821 & 27.427 & 36.075 & 42.065 & 41.379 & \multicolumn{1}{c|}{41.966} & \textbf{72.773} \\
    \textbf{To C++} & 36.647 & 25.350 & 10.110 & 21.112 & 35.228 & 42.364 & 38.929 & \multicolumn{1}{c|}{46.186} & \textbf{74.961} \\
    \textbf{From Python} & 34.137 & 23.232 & 7.898 & 22.327 & 32.635 & 37.693 & 38.455 & \multicolumn{1}{c|}{36.742} & \textbf{64.684} \\
    \textbf{To Python} & 38.058 & 37.066 & 9.783 & 22.059 & 37.976 & 39.044 & 38.263 & \multicolumn{1}{c|}{38.726} & \textbf{75.623} \\
    \textbf{From Java} & 39.101 & 33.405 & 10.935 & 17.493 & 37.552 & 42.000 & 40.209 & \multicolumn{1}{c|}{47.541} & \textbf{76.524} \\
    \textbf{To Java} & 36.113 & 30.714 & 8.760 & 24.077 & 33.058 & 40.349 & 42.851 & \multicolumn{1}{c|}{41.337} & \textbf{63.398} \\
    \textbf{Average} & 36.939 & 31.043 & 9.551 & 22.416 & 35.421 & 40.586 & 40.014 & \multicolumn{1}{c|}{42.083} & \textbf{71.327} \\
    \hline\hline
    \end{tabular}%
    }
  \caption{\textbf{Match-based evaluation} of ExeCoder and baseline models on TransCoder-test-X.}
  \label{tab:baseline_match}%
  \vspace{-0.5cm}
\end{table*}%

\newpage
\section{Implementation Details}

We present detailed implementation details. Deepseek-Coder-6.7b-instruct serves as our basic foundation model. To fine-tune the base model, we employed specific configurations, including a batch size of 128, a sequence length of 3076, a learning rate of 2e-6, a cosine learning rate scheduler, and fp16 mixed precision.

\newpage
\section{Construction of Executability Representation}
\label{sec:graph2text}
In this section, we will present the specific details of encoding syntactic-structure representation and variable-dependency representation, with Figure \ref{fig:g2t_code} illustrating an example of C++ code.

\begin{figure}[htbp]
\centering
\begin{adjustbox}{width=0.3\linewidth,center}
\begin{tabular}{l}

\scriptsize{Input Function}\\
\midrule
\begin{lstlisting}[style=Python, xleftmargin=0cm, basicstyle=\fontencoding{T1}\fontfamily{lmtt}\scriptsize]
int max(int a, int b)
{
    int x = 0;
    if (b > a) 
        x = b;
    else
        x = a;
    return x;
}
\end{lstlisting} \\

 \end{tabular}
 \end{adjustbox}
    \caption{Example C++ code used for encoding executability representations.}
    
\label{fig:g2t_code}
\end{figure}

\newpage
\subsection{Variable-dependency Representation}
\label{sec:dfg_case}
\begin{figure}[htbp]
  \centering
  \includegraphics[width=0.4\linewidth]{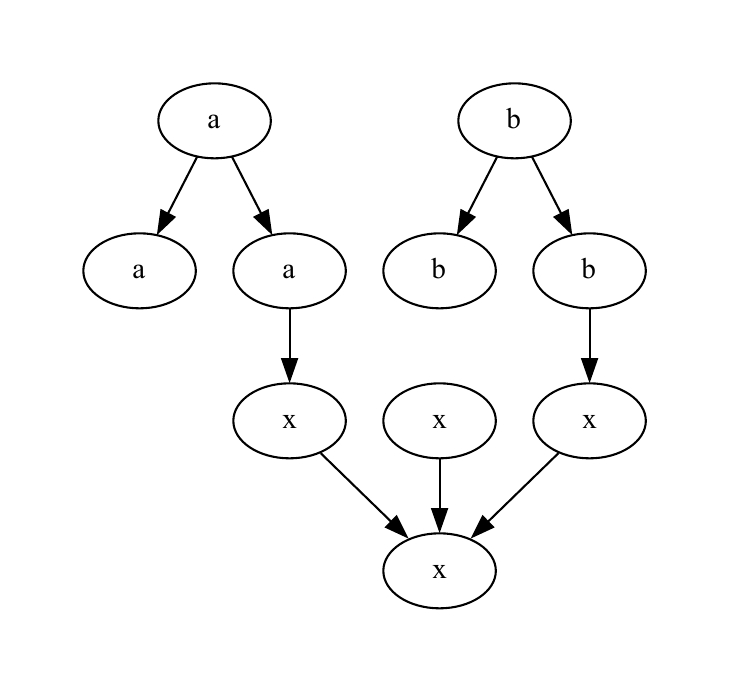}
  \caption{Data flow graph of the example code.}
  \label{fig:dfg_case}
\end{figure}

For variable-dependency representation, ExeCoder first utilizes the code analysis tool tree-sitter to parse its data flow graph (DFG), as illustrated in Figure \ref{fig:dfg_case}. Subsequently, ExeCoder directly assigns a numerical index to each node in the DFG and then uses natural language to represent the node information and edge information of the graph. For the node information, ExeCoder describes the token content of each node. For the edge information, ExeCoder describes the neighboring nodes of each node. The final variable-dependency representation is illustrated in Figure \ref{fig:g2t_dfg}.

\definecolor{grayy}{RGB}{242,242,242}

\lstdefinestyle{lfonts}{
  backgroundcolor=\color{grayy},
  basicstyle   = \footnotesize\ttfamily,
}
\lstdefinestyle{lnumbers}{
}
\lstdefinestyle{llayout}{
  breaklines       = true,
  tabsize          = 2,
  columns          = flexible,
}
\lstdefinestyle{lgeometry}{
}
\lstdefinestyle{lgeneral}{
  style = lfonts,
}
\lstdefinestyle{appendix}{
  language = {Python},
  style    = lgeneral,
}

\lstdefinestyle{lfonts}{
  backgroundcolor=\color{grayy},
  basicstyle   = \footnotesize\ttfamily,
}
\lstdefinestyle{lnumbers}{
}
\lstdefinestyle{llayout}{
  breaklines       = true,
  tabsize          = 2,
  columns          = flexible,
}
\lstdefinestyle{lgeometry}{
}
\lstdefinestyle{lgeneral}{
  style = lfonts,
}
\lstdefinestyle{appendix_cpp}{
  language = {c++},
  style    = lgeneral,
}

\begin{figure*}[htbp]
\centering
\begin{tcolorbox}[width=0.7\textwidth]

\textbf{Variable-dependency Representation}:  
\begin{small}
\begin{lstlisting}[style=appendix, xleftmargin=0cm, basicstyle=\fontencoding{T1}\fontfamily{lmtt}\scriptsize]
G describes a graph among nodes 0, 1, 2, 3, 4, 5, 6, 7, 8, and 9.
In this graph:
Node 0 represents variable a.
Node 1 represents variable b.
Node 2 represents variable x.
Node 3 represents variable b.
Node 4 represents variable a.
Node 5 represents variable x.
Node 6 represents variable b.
Node 7 represents variable x.
Node 8 represents variable a.
Node 9 represents variable x.
In this graph:
Node 3 is connected to node 1.
Node 4 is connected to node 0.
Node 5 is connected to node 6.
Node 6 is connected to node 1.
Node 7 is connected to node 8.
Node 8 is connected to node 0.
Node 9 is connected to nodes 2, 5, 7.
\end{lstlisting}
\end{small}

\end{tcolorbox}
\caption{Variable-dependency representation of the example code.}
\label{fig:g2t_dfg}
\end{figure*}

\newpage
\subsection{Syntactic-structure representation}
\label{sec:ast_case}
\begin{figure*}[htbp]
  \centering
  \includegraphics[width=0.6\linewidth]{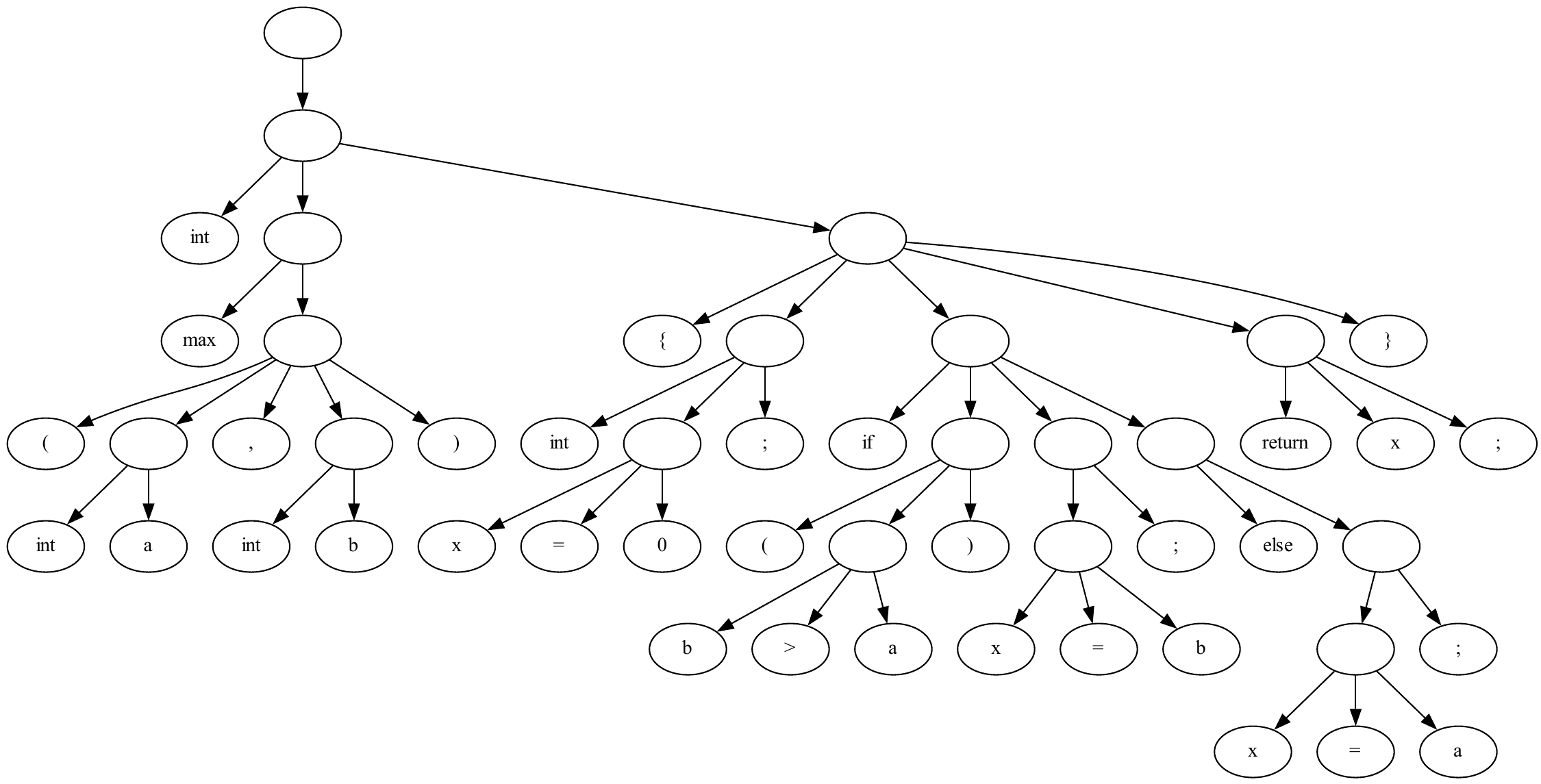}
  \caption{Abstract syntax tree of the example code.}
  \label{fig:ast_case}
  \vspace{-0.5cm}
\end{figure*}

For syntactic-structure representation, ExeCoder first utilizes the code analysis tool tree-sitter to parse its abstract syntax tree (AST) and removes the non-leaf node content, retaining only the tokens at the leaf nodes, as illustrated in Figure \ref{fig:ast_case}. Then, similar to the operations on the DFG, ExeCoder assigns a numerical index to each node of the AST and represents the node and edge information of the graph using natural language. The final syntactic-structure representation is illustrated in Figure \ref{fig:g2t_ast}.

\definecolor{grayy}{RGB}{242,242,242}

\lstdefinestyle{lfonts}{
  backgroundcolor=\color{grayy},
  basicstyle   = \footnotesize\ttfamily,
}
\lstdefinestyle{lnumbers}{
}
\lstdefinestyle{llayout}{
  breaklines       = true,
  tabsize          = 2,
  columns          = flexible,
}
\lstdefinestyle{lgeometry}{
}
\lstdefinestyle{lgeneral}{
  style = lfonts,
}
\lstdefinestyle{appendix}{
  language = {Python},
  style    = lgeneral,
}

\lstdefinestyle{lfonts}{
  backgroundcolor=\color{grayy},
  basicstyle   = \footnotesize\ttfamily,
}
\lstdefinestyle{lnumbers}{
}
\lstdefinestyle{llayout}{
  breaklines       = true,
  tabsize          = 2,
  columns          = flexible,
}
\lstdefinestyle{lgeometry}{
}
\lstdefinestyle{lgeneral}{
  style = lfonts,
}
\lstdefinestyle{appendix_cpp}{
  language = {c++},
  style    = lgeneral,
}

\begin{figure*}[htbp]
\centering
\begin{tcolorbox}[width=0.9\textwidth]

\textbf{Syntactic-structure Representation}:  
\begin{small}
\begin{lstlisting}[style=appendix, xleftmargin=0cm, basicstyle=\fontencoding{T1}\fontfamily{lmtt}\scriptsize]
G describes a graph among nodes 0, 1, 2, 3, 4, 5, 6, 7, 8, 9, 10,
11, 12, 13, 14, 15, 16, 17, 18, 19, 20, 21, 22, 23, 24, 25, 26, 27,
28, 29, 30, 31, 32, 33, 34, 35, 36, 37, 38, 39, 40, 41, 42, 43, 44,
45, 46, 47, 48, 49, 50, and 51.
In this graph:
Node 2 represents code int. Node 5 represents code max.
Node 7 represents code {. Node 11 represents code }.
Node 12 represents code (. Node 14 represents code ,.
Node 16 represents code ). Node 17 represents code int.
Node 19 represents code ;. Node 20 represents code if.
Node 24 represents code return. Node 25 represents code x.
Node 26 represents code ;. Node 27 represents code int.
Node 28 represents code a. Node 29 represents code int.
Node 30 represents code b. Node 31 represents code x.
Node 32 represents code =. Node 33 represents code 0.
Node 34 represents code (. Node 36 represents code ).
Node 38 represents code ;. Node 39 represents code else.
Node 41 represents code b. Node 42 represents code >.
Node 43 represents code a. Node 44 represents code x.
Node 45 represents code =. Node 46 represents code b.
Node 48 represents code ;. Node 49 represents code x.
Node 50 represents code =. Node 51 represents code a.
In this graph:
Node 0 is connected to node 1. Node 1 is connected to nodes 2, 3, 4.
Node 3 is connected to nodes 4, 5. Node 4 is connected to nodes 5, 6, 7, 8, 9.
Node 6 is connected to nodes 7, 8, 9, 10, 11. Node 8 is connected to nodes 9, 10, 11.
Node 9 is connected to nodes 10, 11, 12, 13. Node 10 is connected to nodes 11, 12, 13.
Node 13 is connected to nodes 14, 15. Node 15 is connected to nodes 16, 17.
Node 18 is connected to nodes 19, 20, 21. Node 21 is connected to nodes 22, 23, 24.
Node 22 is connected to nodes 23, 24. Node 23 is connected to nodes 24, 25.
Node 35 is connected to nodes 36, 37, 38. Node 37 is connected to nodes 38, 39, 40.
Node 40 is connected to nodes 41, 42. Node 47 is connected to nodes 48, 49, 50.
\end{lstlisting}
\end{small}

\end{tcolorbox}
\caption{Syntactic-structure representation of the example code.}
\label{fig:g2t_ast}
\end{figure*}

\newpage
\section{Construction of XLCoST-Instruct}
\label{sec:tuning_dataset}
In this section, we will present the construction details of the dataset XLCoST-Instruct, designed for fine-tuning LLMs. XLCoST-Instruct, constructed from the public cross-language code intelligence benchmark dataset XLCoST, aims to enhance the cross-program language understanding capabilities of LLMs by enabling them to learn executability representations of source code through instruction fine-tuning.

\Coder selected parallel data from three programming languages: C++, Python, and Java, to construct a fine-tuning dataset, as these are the most widely used programming languages, and their translation tasks can effectively assess the code translation capabilities of LLMs. Here, the source code refers to complete programs rather than mere code snippets. Subsequently, we encoded three types of executability representations from the source code of XLCoST. For functional semantic representation, we encoded the natural language from the natural language-code pairs provided by XLCoST. For syntactic structure representation and variable dependency representation, we followed the methods outlined in Appendix \ref{sec:graph2text} to encode from the source code. Based on this, a single data instance resulted in four different code representations. Since XLCoST is not specifically designed for instruction fine-tuning of LLMs, \Coder manually constructs instructions tailored for translation tasks, with more detailed information provided in Appendix \ref{sec:tuning_case}.

Finally, we conducted similarity checks and deduplication on the resulting instruction fine-tuning dataset. We employed MinHash with a locality-sensitive hashing (LSH) index to eliminate code instances with a Jaccard similarity greater than 0.85, resulting in approximately 55K instances for the instruction fine-tuning dataset.

\newpage
\section{Example in XLCoST-Instruct}
\label{sec:tuning_case}
In this section, we will present the data instances in the XLCoST-Instruct dataset constructed for fine-tuning LLMs. Figures \ref{fig:tuning_code}, \ref{fig:tuning_nl}, \ref{fig:tuning_ast}, and \ref{fig:tuning_dfg} illustrate the learning of four different representations of the same source code instance in XLCoST-Instruct, namely code semantics, functional-semantic, syntactic-structure, and variable-dependency. 

For the three executability representations, functional-semantic, syntactic-structure, and variable-dependency, ExeCoder utilizes them as auxiliary knowledge during the instruction fine-tuning process. Specifically, ExeCoder concatenates the source code with the corresponding executability representations and then prompts the LLMs to perform code translation tasks based on the respective executability representations, as shown in Figures \ref{fig:tuning_nl}, \ref{fig:tuning_ast}, and \ref{fig:tuning_dfg}.

\definecolor{grayy}{RGB}{242,242,242}

\lstdefinestyle{lfonts}{
  backgroundcolor=\color{grayy},
  basicstyle   = \footnotesize\ttfamily,
}
\lstdefinestyle{lnumbers}{
}
\lstdefinestyle{llayout}{
  breaklines       = true,
  tabsize          = 2,
  columns          = flexible,
}
\lstdefinestyle{lgeometry}{
}
\lstdefinestyle{lgeneral}{
  style = lfonts,
}
\lstdefinestyle{appendix}{
  language = {Python},
  style    = lgeneral,
}

\lstdefinestyle{lfonts}{
  backgroundcolor=\color{grayy},
  basicstyle   = \footnotesize\ttfamily,
}
\lstdefinestyle{lnumbers}{
}
\lstdefinestyle{llayout}{
  breaklines       = true,
  tabsize          = 2,
  columns          = flexible,
}
\lstdefinestyle{lgeometry}{
}
\lstdefinestyle{lgeneral}{
  style = lfonts,
}
\lstdefinestyle{appendix_cpp}{
  language = {c++},
  style    = lgeneral,
}

\begin{figure*}[htbp]
\begin{tcolorbox}

\textbf{Prompt}: Translate the given code from python to cpp. The input Code is marked with \textless Code\textgreater and \textless /Code\textgreater. Please note that the code entered is a complete program with main fuction. \\

\textbf{Input}:  
\begin{small}
\begin{lstlisting}[style=appendix, xleftmargin=0cm, basicstyle=\fontencoding{T1}\fontfamily{lmtt}\scriptsize]
<Code>
def Kmultiples ( n , k ) :
    a = n
    for i in range ( 1 , k + 1 ) :
        print ( "{}*{}={}".format ( n , i , a ) )
        j = 0
        while ( n >= ( 1 << j ) ) :
            a += n & ( 1 << j )
            j += 1
</Code>
\end{lstlisting}
\end{small}

\vspace{0.1cm}

\begin{small}
\textbf{Output}:
\begin{lstlisting}[style=appendix_cpp, xleftmargin=0cm, basicstyle=\fontencoding{T1}\fontfamily{lmtt}\scriptsize]
```cpp
#include <bits/stdc++.h>
using namespace std ;
void Kmultiples ( int n , int k ) {
  int a = n ;
  for ( int i = 1 ;
  i <= k ;
  i ++ ) {
    cout << n << " * " << i << " = " << a << endl ;
    int j = 0 ;
    while ( n >= ( 1 << j ) ) {
      a += n & ( 1 << j ) ;
      j ++ ;
    }
  }
}
```
\end{lstlisting}
\end{small}

\end{tcolorbox}
\caption{Example for learning code semantic in XLCoST-Instruct.}
\label{fig:tuning_code}
\end{figure*}
\definecolor{grayy}{RGB}{242,242,242}

\lstdefinestyle{lfonts}{
  backgroundcolor=\color{grayy},
  basicstyle   = \footnotesize\ttfamily,
}
\lstdefinestyle{lnumbers}{
}
\lstdefinestyle{llayout}{
  breaklines       = true,
  tabsize          = 2,
  columns          = flexible,
}
\lstdefinestyle{lgeometry}{
}
\lstdefinestyle{lgeneral}{
  style = lfonts,
}
\lstdefinestyle{appendix}{
  language = {Python},
  style    = lgeneral,
}

\lstdefinestyle{lfonts}{
  backgroundcolor=\color{grayy},
  basicstyle   = \footnotesize\ttfamily,
}
\lstdefinestyle{lnumbers}{
}
\lstdefinestyle{llayout}{
  breaklines       = true,
  tabsize          = 2,
  columns          = flexible,
}
\lstdefinestyle{lgeometry}{
}
\lstdefinestyle{lgeneral}{
  style = lfonts,
}
\lstdefinestyle{appendix_cpp}{
  language = {c++},
  style    = lgeneral,
}

\begin{figure*}[htbp]
\begin{tcolorbox}

\textbf{Prompt}: Translate the given code from python to cpp. The input contains the source code and a description of the code. The input Code is marked with \textless Code\textgreater and \textless /Code\textgreater. Please note that the code entered is a complete program with main fuction. The description of the code is marked with \textless NL\textgreater and \textless /NL\textgreater. \\ 

\textbf{Input}:  
\begin{small}
\begin{lstlisting}[style=appendix, xleftmargin=0cm, basicstyle=\fontencoding{T1}\fontfamily{lmtt}\scriptsize]
<Code>
def Kmultiples ( n , k ) :
    a = n
    for i in range ( 1 , k + 1 ) :
        print ( "{}*{}={}".format ( n , i , a ) )
        j = 0
        while ( n >= ( 1 << j ) ) :
            a += n & ( 1 << j )
            j += 1
</Code>

<NL>
Generate first K multiples of N ;
Using Bitwise operators | Function to print the first K multiples of N ; 
Print the value of N * i ;
Iterate each bit of N and add pow ( 2 , pos ) , where pos is the index of each set bit ;
Check if current bit at pos j is fixed or not ;
</NL>
\end{lstlisting}
\end{small}

\vspace{0.1cm}

\begin{small}
\textbf{Output}:
\begin{lstlisting}[style=appendix_cpp, xleftmargin=0cm, basicstyle=\fontencoding{T1}\fontfamily{lmtt}\scriptsize]
```cpp
#include <bits/stdc++.h>
using namespace std ;
void Kmultiples ( int n , int k ) {
  int a = n ;
  for ( int i = 1 ;
  i <= k ;
  i ++ ) {
    cout << n << " * " << i << " = " << a << endl ;
    int j = 0 ;
    while ( n >= ( 1 << j ) ) {
      a += n & ( 1 << j ) ;
      j ++ ;
    }
  }
}
```
\end{lstlisting}
\end{small}

\end{tcolorbox}
\caption{Example for learning functional semantic in XLCoST-Instruct.}
\label{fig:tuning_nl}
\end{figure*}
\definecolor{grayy}{RGB}{242,242,242}

\lstdefinestyle{lfonts}{
  backgroundcolor=\color{grayy},
  basicstyle   = \footnotesize\ttfamily,
}
\lstdefinestyle{lnumbers}{
}
\lstdefinestyle{llayout}{
  breaklines       = true,
  tabsize          = 2,
  columns          = flexible,
}
\lstdefinestyle{lgeometry}{
}
\lstdefinestyle{lgeneral}{
  style = lfonts,
}
\lstdefinestyle{appendix}{
  language = {Python},
  style    = lgeneral,
}

\lstdefinestyle{lfonts}{
  backgroundcolor=\color{grayy},
  basicstyle   = \footnotesize\ttfamily,
}
\lstdefinestyle{lnumbers}{
}
\lstdefinestyle{llayout}{
  breaklines       = true,
  tabsize          = 2,
  columns          = flexible,
}
\lstdefinestyle{lgeometry}{
}
\lstdefinestyle{lgeneral}{
  style = lfonts,
}
\lstdefinestyle{appendix_cpp}{
  language = {c++},
  style    = lgeneral,
}

\begin{figure*}[htbp]
\begin{tcolorbox}

\textbf{\small{Prompt}}: \small{Translate the given code from python to cpp. The input contains the source code and a Abstract Syntax Tree of the code. The input Code is marked with \textless Code\textgreater and \textless /Code\textgreater. Please note that the code entered is a complete program with main fuction. The Abstract Syntax Tree of the code is marked with \textless AST\textgreater and \textless /AST\textgreater.} \\ 

\textbf{Input}:  
\begin{small}
\begin{lstlisting}[style=appendix, xleftmargin=0cm, basicstyle=\fontencoding{T1}\fontfamily{lmtt}\tiny]
<Code>
def Kmultiples ( n , k ) :
    a = n
    for i in range ( 1 , k + 1 ) :
        print ( "{}*{}={}".format ( n , i , a ) )
        j = 0
        while ( n >= ( 1 << j ) ) :
            a += n & ( 1 << j )
            j += 1
</Code>
<AST>
G describes a graph among nodes 0, 1, 2, 3, 4, 5, 6, 7, 8, 9, 10, 11, 12,
13, 14, 15, 16, 17, 18, 19, 20, 21, 22, 23, 24, 25, 26, 27, 28, 29, 30, 31,
32, 33, 34, 35, 36, 37, 38, 39, 40, 41, 42, 43, 44, 45, 46, 47, 48, 49, 50,
51, 52, 53, 54, 55, 56, 57, 58, 59, 60, 61, 62, 63, 64, 65, 66, 67, 68, 69,
70, 71, 72, 73, 74, 75, 76, 77, 78, 79, 80, 81, 82, 83, 84, 85, 86, 87, 88,
89, 90, 91, 92, 93, 94, 95, 96, 97, 98, 99, 100, 101, 102, 103, 104, 105, 106,
107, 108, 109, 110, 111, 112, 113, 114, and 115.
In this graph:
Node 5 represents code def. Node 6 represents code Kmultiples. Node 8 represents code :.
Node 13 represents code (. Node 14 represents code n. Node 15 represents code ,.
Node 16 represents code k. Node 17 represents code ). Node 20 represents code N.
Node 21 represents code =. Node 22 represents code 16. Node 23 represents code K.
Node 24 represents code =. Node 25 represents code 7. Node 26 represents code Kmultiples.
Node 29 represents code for. Node 30 represents code i. Node 31 represents code in.
Node 33 represents code :. Node 35 represents code (. Node 36 represents code N.
Node 37 represents code ,. Node 38 represents code K. Node 39 represents code ).
Node 40 represents code a. Node 41 represents code =. Node 42 represents code n.
Node 43 represents code range. Node 48 represents code (. Node 49 represents code 1.
Node 50 represents code ,. Node 52 represents code ). Node 55 represents code while.
Node 57 represents code :. Node 59 represents code k. Node 60 represents code +.
Node 61 represents code 1. Node 62 represents code print. Node 64 represents code j.
Node 65 represents code =. Node 66 represents code 0. Node 67 represents code (.
Node 69 represents code ). Node 72 represents code (. Node 74 represents code ).
Node 75 represents code n. Node 76 represents code >=. Node 82 represents code (.
Node 84 represents code ). Node 85 represents code a. Node 86 represents code +=.
Node 88 represents code j. Node 89 represents code +=. Node 90 represents code 1.
Node 92 represents code .. Node 93 represents code format. Node 94 represents code (.
Node 95 represents code n. Node 96 represents code ,. Node 97 represents code i.
Node 98 represents code ,. Node 99 represents code a. Node 100 represents code ).
Node 101 represents code 1. Node 102 represents code <<. Node 103 represents code j.
Node 104 represents code n. Node 105 represents code &. Node 107 represents code ".
Node 108 represents code {}*{}={}. Node 109 represents code ". Node 110 represents code (.
Node 112 represents code ). Node 113 represents code 1. Node 114 represents code <<.
Node 115 represents code j.
In this graph:
Node 0 is connected to nodes 1, 2, 3, 4. Node 1 is connected to nodes 2, 3, 4, 5, 6.
Node 2 is connected to node 3. Node 3 is connected to node 4.
Node 4 is connected to node 5. Node 7 is connected to nodes 8, 9, 10, 11, 12.
Node 9 is connected to nodes 10, 11. Node 10 is connected to nodes 11, 12, 13.
Node 11 is connected to nodes 12, 13, 14. Node 12 is connected to nodes 13, 14.
Node 18 is connected to node 19. Node 19 is connected to nodes 20, 21, 22, 23, 24, 25.
Node 27 is connected to nodes 28, 29, 30, 31, 32. Node 28 is connected to nodes 29, 30, 31.
Node 32 is connected to nodes 33, 34. Node 34 is connected to nodes 35, 36, 37.
Node 44 is connected to nodes 45, 46, 47, 48, 49. Node 45 is connected to node 46.
Node 46 is connected to node 47. Node 47 is connected to nodes 48, 49, 50, 51.
Node 51 is connected to nodes 52, 53, 54. Node 53 is connected to nodes 54, 55.
Node 54 is connected to nodes 55, 56, 57. Node 56 is connected to nodes 57, 58, 59.
Node 58 is connected to nodes 59, 60. Node 63 is connected to nodes 64, 65, 66.
Node 68 is connected to nodes 69, 70, 71. Node 70 is connected to node 71.
Node 71 is connected to node 72. Node 73 is connected to nodes 74, 75.
Node 77 is connected to nodes 78, 79, 80. Node 78 is connected to nodes 79, 80, 81.
Node 79 is connected to nodes 80, 81, 82. Node 80 is connected to nodes 81, 82, 83.
Node 81 is connected to nodes 82, 83, 84, 85, 86, 87, 88. Node 83 is connected to nodes 84, 85, 86.
Node 87 is connected to nodes 88, 89, 90. Node 91 is connected to nodes 92, 93, 94.
Node 106 is connected to nodes 107, 108, 109. Node 111 is connected to nodes 112, 113, 114.
</AST>
\end{lstlisting}
\end{small}


\begin{small}
\textbf{Output}:
\begin{lstlisting}[style=appendix_cpp, xleftmargin=0cm, basicstyle=\fontencoding{T1}\fontfamily{lmtt}\scriptsize]
```cpp
#include <bits/stdc++.h>
using namespace std ;
void Kmultiples ( int n , int k ) {
  int a = n ;
  for ( int i = 1 ; i <= k ; i ++ ) {
    cout << n << " * " << i << " = " << a << endl ;
    int j = 0 ;
    while ( n >= ( 1 << j ) ) {
      a += n & ( 1 << j ) ;
      j ++ ;}}}
```
\end{lstlisting}
\end{small}

\end{tcolorbox}
\caption{Example for learning syntactic structure in XLCoST-Instruct.}
\label{fig:tuning_ast}
\end{figure*}
\definecolor{grayy}{RGB}{242,242,242}

\lstdefinestyle{lfonts}{
  backgroundcolor=\color{grayy},
  basicstyle   = \footnotesize\ttfamily,
}
\lstdefinestyle{lnumbers}{
}
\lstdefinestyle{llayout}{
  breaklines       = true,
  tabsize          = 2,
  columns          = flexible,
}
\lstdefinestyle{lgeometry}{
}
\lstdefinestyle{lgeneral}{
  style = lfonts,
}
\lstdefinestyle{appendix}{
  language = {Python},
  style    = lgeneral,
}

\lstdefinestyle{lfonts}{
  backgroundcolor=\color{grayy},
  basicstyle   = \footnotesize\ttfamily,
}
\lstdefinestyle{lnumbers}{
}
\lstdefinestyle{llayout}{
  breaklines       = true,
  tabsize          = 2,
  columns          = flexible,
}
\lstdefinestyle{lgeometry}{
}
\lstdefinestyle{lgeneral}{
  style = lfonts,
}
\lstdefinestyle{appendix_cpp}{
  language = {c++},
  style    = lgeneral,
}

\begin{figure*}[htbp]
\begin{tcolorbox}

\textbf{Prompt}: Translate the given code from python to cpp. The input contains the source code and a Dataflow Graph of the code. The input Code is marked with \textless Code\textgreater and \textless /Code\textgreater. Please note that the code entered is a complete program with main fuction. The Dataflow Graph of the code is marked with \textless DFG\textgreater and \textless /DFG\textgreater. \\ 

\textbf{Input}:  
\begin{small}
\begin{lstlisting}[style=appendix, xleftmargin=0cm, basicstyle=\fontencoding{T1}\fontfamily{lmtt}\scriptsize]
<Code>
def Kmultiples ( n , k ) :
    a = n
    for i in range ( 1 , k + 1 ) :
        print ( "{}*{}={}".format ( n , i , a ) )
        j = 0
        while ( n >= ( 1 << j ) ) :
            a += n & ( 1 << j )
            j += 1
</Code>

<DFG>
G describes a graph among nodes 0, 1, 2, 3, 4, 5, 6, 7, 8, 9, 10,
11, 12, 13, 14, 15, 16, 17, 18, 19, 20, 21, 22, 23, 24, 25, 26, 27,
28, 29, 30, and 31.
In this graph:
Node 0 represents variable Kmultiples. Node 1 represents variable n.
Node 2 represents variable k. Node 3 represents variable a.
Node 4 represents variable n. Node 5 represents variable i.
Node 6 represents variable range. Node 7 represents variable 1.
Node 8 represents variable k. Node 9 represents variable 1.
Node 10 represents variable print. Node 11 represents variable format.
Node 12 represents variable n. Node 13 represents variable i.
Node 14 represents variable a. Node 15 represents variable j.
Node 16 represents variable 0. Node 17 represents variable n.
Node 18 represents variable j. Node 19 represents variable a.
Node 20 represents variable n. Node 21 represents variable 1.
Node 22 represents variable j. Node 23 represents variable j.
Node 24 represents variable 1. Node 25 represents variable N.
Node 26 represents variable 16. Node 27 represents variable K.
Node 28 represents variable 7. Node 29 represents variable Kmultiples.
Node 30 represents variable N. Node 31 represents variable K.
In this graph:
Node 3 is connected to node 4. Node 4 is connected to node 1.
Node 5 is connected to nodes 6, 7, 8, 9. Node 6 is connected to node 6.
Node 8 is connected to node 2. Node 10 is connected to node 10.
Node 11 is connected to node 11. Node 12 is connected to node 1.
Node 13 is connected to node 5. Node 14 is connected to nodes 3, 19.
Node 15 is connected to node 16. Node 17 is connected to node 1.
Node 18 is connected to nodes 15, 23. Node 19 is connected to nodes 20, 21, 22.
Node 20 is connected to node 1. Node 22 is connected to nodes 15, 23.
Node 23 is connected to node 24. Node 25 is connected to node 26.
Node 27 is connected to node 28. Node 29 is connected to node 0.
Node 30 is connected to node 25. Node 31 is connected to node 27.
</DFG>
\end{lstlisting}
\end{small}

\vspace{0.1cm}

\begin{small}
\textbf{Output}:
\begin{lstlisting}[style=appendix_cpp, xleftmargin=0cm, basicstyle=\fontencoding{T1}\fontfamily{lmtt}\scriptsize]
```cpp
#include <bits/stdc++.h>
using namespace std ;
void Kmultiples ( int n , int k ) {
  int a = n ;
  for ( int i = 1 ;
  i <= k ;
  i ++ ) {
    cout << n << " * " << i << " = " << a << endl ;
    int j = 0 ;
    while ( n >= ( 1 << j ) ) {
      a += n & ( 1 << j ) ;
      j ++ ;
    }
  }
}
```
\end{lstlisting}
\end{small}

\end{tcolorbox}
\caption{Example for learning variable dependency in XLCoST-Instruct.}
\label{fig:tuning_dfg}
\end{figure*}

\newpage
\section{Construction of TransCoder-test-X}
\label{sec:TransCoder-test-X}
In this section, we will present the detailed process of enhancing TransCoder-test to obtain TransCoder-test-X. 

We first manually added various parameter passing methods under equivalent implementations to the unit test templates, as illustrated in Figures \ref{fig:evaluation_cpp_original} and \ref{fig:evaluation_cpp_enhanced}. Specifically, Figure \ref{fig:evaluation_cpp_original} depicts the unit test template in TransCoder-test, which sets specific parameter passing methods for array-type variables in the main function, passing parameters from param0 to the generated function for unit testing. However, this parameter passing method only allows parameters to be passed for array-type variables; when functions are implemented using vector containers, even if the generated function is functionally equivalent to the reference function, the predefined unit test parameters cannot be passed, and the unit tests cannot be executed as expected. To address this, our solution was to manually add various parameter passing methods for equivalent implementations to the unit test templates. Specifically, we manually modified the parameter passing methods in the main function and added template wrapper functions, as shown in Figure \ref{fig:evaluation_cpp_enhanced}. The template wrapper function determines the method of parameter passing by assessing the implementation of the function.

We also manually aligned the return types of the parallel data. During the evaluation of TransCoder-test, we found that some return types of the parallel data were misaligned, as shown in Figure \ref{fig:evaluation_java}. Figure \ref{fig:evaluation_java} illustrates a translation from C++ to Java, where the return type of the source function is int and the return type of the reference function is boolean. The correct translation of the source function should yield a return type of int, rather than the return type of boolean from the reference function. However, in Java, int and boolean types cannot be directly equated, which means that even if the translation is correct, the unit tests cannot execute as intended. To address this issue, we manually aligned the return types of the parallel data. 

Furthermore, we corrected inherent errors in the unit test templates present in TransCoder-test, as shown in Figure \ref{fig:evaluation_template}. In Figure \ref{fig:evaluation_template}, the original code in TransCoder-test contains an extraneous comma in the parameters of the main function, preventing it from compiling correctly, which we manually corrected. 

Finally, the processed test set is referred to as TransCoder-test-X, where the unit tests are capable of assessing the equivalence of code with different implementation methods and evaluating LLMs' code translation abilities.

\newpage
\definecolor{grayy}{RGB}{242,242,242}

\lstdefinestyle{lfonts}{
  backgroundcolor=\color{grayy},
  basicstyle   = \footnotesize\ttfamily,
}
\lstdefinestyle{lnumbers}{
}
\lstdefinestyle{llayout}{
  breaklines       = true,
  tabsize          = 2,
  columns          = flexible,
}
\lstdefinestyle{lgeometry}{
}
\lstdefinestyle{lgeneral}{
  style = lfonts,
}
\lstdefinestyle{appendix_cpp}{
  language = {c++},
  style    = lgeneral,
}

\begin{figure*}[htbp]
\begin{tcolorbox} 
\begin{small}
\begin{lstlisting}[style=appendix_cpp, xleftmargin=0cm, basicstyle=\fontencoding{T1}\fontfamily{lmtt}\scriptsize]
#include <iostream>
#include <cstdlib>
#include <string>
#include <vector>
#include <fstream>
#include <iomanip>
#include <bits/stdc++.h>
using namespace std;
int f_gold ( int arr [ ], int n, int x ) {
  int i;
  for ( i = 0;
  i < n;
  i ++ ) {
    if ( arr [ i ] == x ) return i;
  }
  return - 1;
}


//TOFILL

int main() {
    int n_success = 0;
    vector<vector<int>> param0 
    {{4,5,5,11,13,14,15,19,22,22,23,26,29,29,36,44,48,49,65,65,
    67,68,70,76,79,79,81,85,88,91,91,92,92,97},
    
    {-24,-78,-32,-48,0,4,-42},
    {0,0,0,0,0,0,0,1,1,1,1},
    {38,14,75,16,91,11,98,43,67,9,21,10,82,72,32,81,48,60,2,91,10,90,12,83},
    
    {-92,-92,-82,-80,-76,-66,-64,-64,-56,-48,-38,-38,-34,-32,-32,-10,
    -8,-6,-2,0,8,10,18,20,22,22,30,34,38,38,38,44,50,52,56,64,64,66,70,76,88},
    {0,1,1,0,0,1,1,0,0,0,1,1,1,1},
    
    {1,4,4,4,4,8,12,13,14,14,22,25,25,27,29,33,36,38,40,40,40,
    41,47,47,47,48,48,50,51,52,52,52,55,56,59,59,62,64,66,77,82,84,90,91,91,93},
    
    {-90,-60,-58,-72,92,54,-32,-70,-94,18,64,-90,-90,-56,82,-14,-74,-96,-90,
    -8,-48,76,-28,10,-52,-8,-46,-32,82,46,58,92,4,48,-96,-66,60,60,62,-68},
    
    {0,0,0,0,0,0,1,1,1,1},{42,17,77,96,72,36,74,97,7,94,80,7,27,58,49,81,51,9}};
    
    vector<int> param1 {17,4,6,17,25,11,38,22,8,16};
    vector<int> param2 {5,0,0,75,25,-1,4,22,8,11};
    for(int i = 0; i < param0.size(); ++i)
    {
        if(search(&param0[i].front(),param1[i],param2[i]) == 
        f_gold(&param0[i].front(),param1[i],param2[i]))
        {
            n_success+=1;
        }
    }
    cout << "#Results:" << " " << n_success << ", " << param0.size();
    return 0;
}

\end{lstlisting}
\end{small}

\end{tcolorbox}
\caption{Original code in TransCoder-test, evaluating specific code implementations.}
\label{fig:evaluation_cpp_original}
\end{figure*}

\newpage
\definecolor{grayy}{RGB}{242,242,242}

\lstdefinestyle{lfonts}{
  backgroundcolor=\color{grayy},
  basicstyle   = \footnotesize\ttfamily,
}
\lstdefinestyle{lnumbers}{
}
\lstdefinestyle{llayout}{
  breaklines       = true,
  tabsize          = 2,
  columns          = flexible,
}
\lstdefinestyle{lgeometry}{
}
\lstdefinestyle{lgeneral}{
  style = lfonts,
}
\lstdefinestyle{appendix_cpp}{
  language = {c++},
  style    = lgeneral,
}

\begin{figure*}[htbp]
\begin{tcolorbox} 
\begin{small}
\begin{lstlisting}[style=appendix_cpp, xleftmargin=0cm, basicstyle=\fontencoding{T1}\fontfamily{lmtt}\scriptsize]
#include <iostream>
#include <cstdlib>
#include <string>
#include <vector>
#include <fstream>
#include <iomanip>
#include <bits/stdc++.h>
using namespace std;

int f_gold ( int arr [ ], int n, int x ) {
    int i;
    for ( i = 0;
    i < n;
    i ++ ) {
        if ( arr [ i ] == x ) return i;
    }
    return - 1;
}

//TOFILL

template <typename T>
int f_gold(T arr, int n, int x) {
    if constexpr (is_same_v<T, vector<int>>) {
        return f_gold(&arr.front(), n, x);
    } else {
        return f_gold(arr, n, x);
    }
}

template <typename T>
int search(T arr, int n, int x) {
    if constexpr (is_same_v<T, vector<int>>) {
        return search(&arr.front(), n, x);
    } else {
        return search(arr, n, x);
    }
}

int main() {
    int n_success = 0;
    vector<vector<int>> param0 
    {{4,5,5,11,13,14,15,19,22,22,23,26,29,29,36,44,48,49,65,65,
    67,68,70,76,79,79,81,85,88,91,91,92,92,97},
    
    {-24,-78,-32,-48,0,4,-42},
    {0,0,0,0,0,0,0,1,1,1,1},
    {38,14,75,16,91,11,98,43,67,9,21,10,82,72,32,81,48,60,2,91,10,90,12,83},
    
    {-92,-92,-82,-80,-76,-66,-64,-64,-56,-48,-38,-38,-34,-32,-32,-10,
    -8,-6,-2,0,8,10,18,20,22,22,30,34,38,38,38,44,50,52,56,64,64,66,70,76,88},
    {0,1,1,0,0,1,1,0,0,0,1,1,1,1},
    
    {1,4,4,4,4,8,12,13,14,14,22,25,25,27,29,33,36,38,40,40,40,
    41,47,47,47,48,48,50,51,52,52,52,55,56,59,59,62,64,66,77,82,84,90,91,91,93},
    
    {-90,-60,-58,-72,92,54,-32,-70,-94,18,64,-90,-90,-56,82,-14,-74,-96,-90,
    -8,-48,76,-28,10,-52,-8,-46,-32,82,46,58,92,4,48,-96,-66,60,60,62,-68},
    
    {0,0,0,0,0,0,1,1,1,1},{42,17,77,96,72,36,74,97,7,94,80,7,27,58,49,81,51,9}};
    
    vector<int> param1 {17,4,6,17,25,11,38,22,8,16};
    vector<int> param2 {5,0,0,75,25,-1,4,22,8,11};
    for(int i = 0; i < param0.size(); ++i)
    {
        if(search(param0[i], param1[i], param2[i]) == f_gold(param0[i], param1[i], param2[i]))
        {
            n_success+=1;
        }
    }
    cout << "#Results:" << " " << n_success << ", " << param0.size();
    return 0;
}

\end{lstlisting}
\end{small}

\end{tcolorbox}
\caption{Enhanced code in TransCoder-test-X, evaluating various equivalent code implementations.}
\label{fig:evaluation_cpp_enhanced}
\end{figure*}

\newpage
\definecolor{grayy}{RGB}{242,242,242}

\lstdefinestyle{lfonts}{
  backgroundcolor=\color{grayy},
  basicstyle   = \footnotesize\ttfamily,
}
\lstdefinestyle{lnumbers}{
}
\lstdefinestyle{llayout}{
  breaklines       = true,
  tabsize          = 2,
  columns          = flexible,
}
\lstdefinestyle{lgeometry}{
}
\lstdefinestyle{lgeneral}{
  style = lfonts,
}
\lstdefinestyle{appendix}{
  language = {Python},
  style    = lgeneral,
}

\lstdefinestyle{lfonts}{
  backgroundcolor=\color{grayy},
  basicstyle   = \footnotesize\ttfamily,
}
\lstdefinestyle{lnumbers}{
}
\lstdefinestyle{llayout}{
  breaklines       = true,
  tabsize          = 2,
  columns          = flexible,
}
\lstdefinestyle{lgeometry}{
}
\lstdefinestyle{lgeneral}{
  style = lfonts,
}
\lstdefinestyle{appendix_cpp}{
  language = {c++},
  style    = lgeneral,
}

\begin{figure*}[htbp]
\begin{tcolorbox}

\textbf{C++ Code}:  
\begin{small}
\begin{lstlisting}[style=appendix_cpp, xleftmargin=0cm, basicstyle=\fontencoding{T1}\fontfamily{lmtt}\scriptsize]
int f_gold ( int num ) {
  if ( num < 0 ) return f_gold ( - num );
  if ( num == 0 || num == 7 ) return 1;
  if ( num < 10 ) return 0;
  return f_gold ( num / 10 - 2 * ( num - num / 10 * 10 ) );
}
\end{lstlisting}
\end{small}

\vspace{0.1cm}

\begin{small}
\textbf{Original Java Code}:
\begin{lstlisting}[style=appendix_cpp, xleftmargin=0cm, basicstyle=\fontencoding{T1}\fontfamily{lmtt}\scriptsize]
static boolean f_gold ( int num ) {
  if ( num < 0 ) return f_gold ( - num ) ;
  if ( num == 0 || num == 7 ) return true ;
  if ( num < 10 ) return false ;
  return f_gold ( num / 10 - 2 * ( num - num / 10 * 10 ) ) ;
}
\end{lstlisting}
\end{small}

\vspace{0.1cm}

\begin{small}
\textbf{Enhanced Java Code}:
\begin{lstlisting}[style=appendix_cpp, xleftmargin=0cm, basicstyle=\fontencoding{T1}\fontfamily{lmtt}\scriptsize]
static int f_gold ( int num ) {
  if ( num < 0 ) return f_gold ( - num ) ;
  if ( num == 0 || num == 7 ) return 1 ;
  if ( num < 10 ) return 0 ;
  return f_gold ( num / 10 - 2 * ( num - num / 10 * 10 ) ) ;
}
\end{lstlisting}
\end{small}

\end{tcolorbox}
\caption{Alignment of return types for parallel data in TransCoder-test.}
\label{fig:evaluation_java}
\end{figure*}

\newpage
\definecolor{grayy}{RGB}{242,242,242}

\lstdefinestyle{lfonts}{
  backgroundcolor=\color{grayy},
  basicstyle   = \footnotesize\ttfamily,
}
\lstdefinestyle{lnumbers}{
}
\lstdefinestyle{llayout}{
  breaklines       = true,
  tabsize          = 2,
  columns          = flexible,
}
\lstdefinestyle{lgeometry}{
}
\lstdefinestyle{lgeneral}{
  style = lfonts,
}
\lstdefinestyle{appendix}{
  language = {Python},
  style    = lgeneral,
}

\lstdefinestyle{lfonts}{
  backgroundcolor=\color{grayy},
  basicstyle   = \footnotesize\ttfamily,
}
\lstdefinestyle{lnumbers}{
}
\lstdefinestyle{llayout}{
  breaklines       = true,
  tabsize          = 2,
  columns          = flexible,
}
\lstdefinestyle{lgeometry}{
}
\lstdefinestyle{lgeneral}{
  style = lfonts,
}
\lstdefinestyle{appendix_cpp}{
  language = {c++},
  style    = lgeneral,
}

\begin{figure*}[htbp]
\begin{tcolorbox}

\textbf{Original Evaluation Code}:  
\begin{small}
\begin{lstlisting}[style=appendix, xleftmargin=0cm, basicstyle=\fontencoding{T1}\fontfamily{lmtt}\scriptsize]
if __name__ == '__main__':
    param = [
    ([6,7,15,42,47,54,56,59,59,64,68,70,71,75,91,93], 0, 15, 71),
([6,7,15,42,47,56,54,59,59,64,68,71,70, 75,91,93], 0, 15,    71),
([-92,-96,-68,-40,70], 0, 4, ,    -96),
([-92,-86,-68,-40,70], 0, 4,    20),
([-3,-1,0,30,10,45,70,60], 0, 7,    0),
([-3,-1,0,10,5,45,60,50], 0, 7,    12),
([-3,-1,0,10,30,45,60,70], 0, 7,    18),
([0,0,1], 0, 2,    20),
([1,1,1], 0, 2,    17),
([30,2,30,45], 0, 3,    28)
        ]
    n_success = 0
    for i, parameters_set in enumerate(param):
        if binarySearch(*parameters_set) == f_gold(*parameters_set):
            n_success+=1
    print("#Results: %i, %i" % (n_success, len(param)))
\end{lstlisting}
\end{small}

\vspace{0.1cm}

\begin{small}
\textbf{Enhanced Evaluaion Code}:
\begin{lstlisting}[style=appendix, xleftmargin=0cm, basicstyle=\fontencoding{T1}\fontfamily{lmtt}\scriptsize]
if __name__ == '__main__':
    param = [
    ([6,7,15,42,47,54,56,59,59,64,68,70,71,75,91,93], 0, 15, 71),
([6,7,15,42,47,56,54,59,59,64,68,71,70, 75,91,93], 0, 15,    71),
([-92,-96,-68,-40,70], 0, 4,    -96),
([-92,-86,-68,-40,70], 0, 4,    20),
([-3,-1,0,30,10,45,70,60], 0, 7,    0),
([-3,-1,0,10,5,45,60,50], 0, 7,    12),
([-3,-1,0,10,30,45,60,70], 0, 7,    18),
([0,0,1], 0, 2,    20),
([1,1,1], 0, 2,    17),
([30,2,30,45], 0, 3,    28)
        ]
    n_success = 0
    for i, parameters_set in enumerate(param):
        if binarySearch(*parameters_set) == f_gold(*parameters_set):
            n_success+=1
    print("#Results: %i, %i" % (n_success, len(param)))
\end{lstlisting}
\end{small}

\end{tcolorbox}
\caption{Inherent errors in the unit test template present in TransCoder-test.}
\label{fig:evaluation_template}
\end{figure*}

\clearpage
\section{Example in TransCoder-test-X}
In this section, we will present the evaluation instance in TransCoder-test-X and example of ExeCoder's responses to the instance, as shown in Figure \ref{fig:evaluation_case}. In this example, ExeCoder performs the translation from C++ to Python, where the function calculates the number of unique non-zero elements in a given matrix. 
 
\definecolor{grayy}{RGB}{242,242,242}

\lstdefinestyle{lfonts}{
  backgroundcolor=\color{grayy},
  basicstyle   = \footnotesize\ttfamily,
}
\lstdefinestyle{lnumbers}{
}
\lstdefinestyle{llayout}{
  breaklines       = true,
  tabsize          = 2,
  columns          = flexible,
}
\lstdefinestyle{lgeometry}{
}
\lstdefinestyle{lgeneral}{
  style = lfonts,
}
\lstdefinestyle{appendix}{
  language = {Python},
  style    = lgeneral,
}

\begin{figure*}[htbp]
\begin{tcolorbox}

\textbf{Prompt}: Translate the given code from cpp to python. The input Code is marked with \textless Code\textgreater and \textless/Code\textgreater. \\ 

\textbf{Input}:  
\begin{small}
\begin{lstlisting}[style=appendix, xleftmargin=0cm, basicstyle=\fontencoding{T1}\fontfamily{lmtt}\scriptsize]
<Code>
int countUnique ( int mat [ ] [ MAX ] , int n , int m ) {
  int rowsum [ n ] , colsum [ m ] ;
  memset ( colsum , 0 , sizeof ( colsum ) ) ;
  memset ( rowsum , 0 , sizeof ( rowsum ) ) ;
  for ( int i = 0 ;
  i < n ;
  i ++ ) for ( int j = 0 ;
  j < m ;
  j ++ ) if ( mat [ i ] [ j ] ) {
    rowsum [ i ] ++ ;
    colsum [ j ] ++ ;
  }
  int uniquecount = 0 ;
  for ( int i = 0 ;
  i < n ;
  i ++ ) for ( int j = 0 ;
  j < m ;
  j ++ ) if ( mat [ i ] [ j ] && rowsum [ i ] == 1 && colsum [ j ] == 1 ) uniquecount ++ ;
  return uniquecount ;
}
</Code>
\end{lstlisting}
\end{small}

\vspace{0.1cm}

\begin{small}
\textbf{ExeCoder Response}:
\begin{lstlisting}[style=appendix, xleftmargin=0cm, basicstyle=\fontencoding{T1}\fontfamily{lmtt}\scriptsize]
```python
def countUnique ( mat , n , m ) :
    rowsum = [ 0 ] * n
    colsum = [ 0 ] * m
    for i in range ( n ) :
        for j in range ( m ) :
            if ( mat [ i ] [ j ] ) :
                rowsum [ i ] += 1
                colsum [ j ] += 1
    uniquecount = 0
    for i in range ( n ) :
        for j in range ( m ) :
            if ( mat [ i ] [ j ] and rowsum [ i ] == 1 and colsum [ j ] == 1 ) :
                uniquecount += 1
    return uniquecount
```
\end{lstlisting}
\end{small}

\end{tcolorbox}
\caption{Example of ExeCoder's evaluation on TransCoder-test-X.}
\label{fig:evaluation_case}
\end{figure*}

\end{document}